\documentclass[preprint2]{aastex}
\usepackage{amsmath,amstext}
\usepackage{tikz,xspace}
\usepackage{fancyvrb}
\usetikzlibrary{shapes.geometric,arrows,positioning,calc}
\usepackage[breaklinks,colorlinks,citecolor=blue,linkcolor=magenta]{hyperref}
\usepackage[all]{hypcap} 

\newcommand\json{{\tt JSON}\xspace}
\newcommand\git{{\tt git}\xspace}
\newcommand\xml{{\tt XML}\xspace}
\newcommand\dt{{\tt DataTables}\xspace}
\newcommand\github{{\tt github}\xspace}
\newcommand\python{{\tt Python}\xspace}
\newcommand\osc{{\it Open Supernova Catalog}\xspace}
\newcommand\astrocats{{\tt AstroCats}\xspace}

\tikzstyle{user} = [ellipse, fill=yellow!20, text width=1.5cm, minimum width=2cm, text centered, draw=black, minimum height=1cm]
\tikzstyle{file} = [rectangle, rounded corners, text width=1.5cm, minimum width=2cm, minimum height=1cm, text centered, draw=black, fill=red!30]
\tikzstyle{web} = [trapezium, trapezium left angle=80, text width=1.5cm, trapezium stretches body, trapezium right angle=100, text centered, draw=black, fill=blue!30]
\tikzstyle{process} = [rectangle, text width=1.5cm, minimum width=2cm, minimum height=1cm, text centered, draw=black, fill=orange!30]
\tikzstyle{repo} = [diamond, text width=1cm, minimum width=1cm, minimum height=1cm, text centered, draw=black, fill=green!30]
\tikzstyle{arrow} = [thick,->,>=stealth]

\shorttitle{Open Supernova Catalog}
\shortauthors{Guillochon et al.}

\begin{document}

\nocite{Matheson00b,Matheson00a} \nocite{Taddia1206} \nocite{Taddia1206} \nocite{Nicholl14} \nocite{Cao15}

\title{An Open Catalog for Supernova Data}
\author{James Guillochon\altaffilmark{1}, Jerod Parrent\altaffilmark{1}, Luke Zoltan Kelley\altaffilmark{1}, and Raffaella Margutti\altaffilmark{2,3}}
\affil{$^1$Harvard-Smithsonian Center for Astrophysics, 60 Garden St., Cambridge, MA 02138, USA}
\affil{$^2$Center for Interdisciplinary Exploration and Research in Astrophysics (CIERA) and Department of Physics and Astrophysics, Northwestern University, Evanston, IL 60208}
\affil{$^3$New York University, Physics department, 4 Washington Place, New York, NY 10003, USA}
\email{jguillochon@cfa.harvard.edu}

\begin{abstract}
We present the \osc, an online collection of observations and metadata for presently 36,000+ supernovae and related candidates. The catalog is freely available on the web \mbox{(\url{https://sne.space})}, with its main interface having been designed to be a user-friendly, rapidly-searchable table accessible on desktop and mobile devices. In addition to the primary catalog table containing supernova metadata, an individual page is generated for each supernova which displays its available metadata, light curves, and spectra spanning X-ray to radio frequencies. The data presented in the catalog is automatically rebuilt on a daily basis and is constructed by parsing several dozen sources, including the data presented in the supernova literature and from secondary sources such as other web-based catalogs. Individual supernova data is stored in the hierarchical, human- and machine-readable \json format, with the entirety of each supernova's data being contained within a single \json file bearing its name. The setup we present here, which is based upon open source software maintained via \git repositories hosted on \github, enables anyone to download the entirety of the supernova dataset to their home computer in minutes, and to make contributions of their own data back to the catalog via \git. As the supernova dataset continues to grow, especially in the upcoming era of all-sky synoptic telescopes which will increase the total number of events by orders of magnitude, we hope that the catalog we have designed will be a valuable tool for the community to analyze both historical and contemporary supernovae.
\end{abstract}

\maketitle

\section{Introduction}

Whether observed or synthesized, data begets analysis, and organized and open access to this data for independent analyses or large case studies is critical to its utility. For relatively young sub-fields in astronomy, such as the exoplanet community, many have advocated for open data access \citep{Wright:2011a,Rein12}, and the result has been an abundance of rapid scientific progress enabled not just by the data collectors, but also the broader community with access to that data. Older astronomical sub-fields such as variable stars have similarly benefited from communal efforts, including support from secondary schools and amateur astronomers \citep{Kinne12}.

For discoveries of exploded stars and their remnants, there have been several historical catalogs that detail all available metadata \citep{Clark76,FLIN79,Green84,Green88,Barbon99,Tsvetkov04,Lennarz12}. Today resources on the web, such as David Bishop's
``Latest Supernova'' page and Dave Green's ``Catalog of Supernova Remnants'' have since
facilitated the continuous influx of more recent discoveries.\footnote{\url{http://www.rochesterastronomy.org/supernova.html}}$^{,}$\footnote{\url{http://www.mrao.cam.ac.uk/surveys/snrs/}} 

As for supernova science and the collection of published observations, it is often said the data scene is as messy as it is unpopular. Foreseeing the need for an extensive comparative study of all types of supernovae, David Branch and associates issued a bulletin in 1982 for the collection of all spectroscopic and photometric supernova data published within the literature. In addition to constructing an interactive database, their vision included displaying the data graphically in a uniform format to facilitate comparative studies of supernovae \citep{Branch82}.

Later during the 1990s, David Jeffery kept a personal repository of available data\footnote{\url{https://www.nhn.ou.edu/~jeffery/astro/sne/spectra/spectra.html}}. This effort eventually led to the creation of the Supernova Spectrum archive (SuSpect) by the Supernova Group at the University of Oklahoma \citep{Richardson01}, which collected nearly 1800 spectra by 2007 through donations, individual requests, and digitizing in some instances \citep{Casebeer98}. A unified collection of photometric observations was also pursued by SuSpect, and to a larger extent by the ITEP-SAI Supernova Light Curve Catalog\footnote{\url{http://dau.itep.ru/sn/lc}} built by P.V.Baklanov, S.I.Blinnikov, D.Yu.Tsvetkov, and N.N.Pavlyuk. 

Additional repositories of supernova observations have since included the CfA SN Archive\footnote{\url{https://www.cfa.harvard.edu/supernova/SNarchive.html}}, UC SuperNova DataBase\footnote{\url{http://heracles.astro.berkeley.edu/sndb/}} \citep{Silverman12a}, and select data have also been released by: the Nearby SuperNova factory (SNfactory, \citealt{Aldering02}); the Supernova Legacy Survey (SNLS, \citealt{Balland09}); the Palomar Transient Factory (PTF, \citealt{Rau09}); the La Silla-QUEST Southern Hemisphere Variability Survey (LSQ, \citealt{Baltay12}); the Carnegie Supernova Project (CSP, \citealt{Folatelli13}); the All-Sky Automated Survey for Supernovae (ASAS-SN, \citealt{Shappee14}); the Panoramic Survey Telescope \& Rapid Response System (PS1, \citealt{Rest14}); and the Public ESO Spectroscopic Survey of Transient Objects (PESSTO, \citealt{Smartt15PESSTO}). Since the more recent creation of the Weizmann Interactive Supernova data REPository\footnote{\url{http://wiserep.weizmann.ac.il/}} (WISeREP, \citealt{WISEREP}) and the Transient Name Server\footnote{\url{https://wis-tns.weizmann.ac.il/}}, there has been an ever growing interest in bringing supernova science closer to an ``era of big data'' with the collection of over 13,000 publicly available supernova spectra.

Still, no single repository has amassed all available lights curves and spectra, and numerous published datasets remain bounded to the original paper. Furthermore, no repository currently exists for X-ray and radio observations of supernovae. Stephan Immler once maintained a webpage with the names of supernovae detected in the X-rays, however this site\footnote{\url{http://lheawww.gsfc.nasa.gov/users/immler/supernovae_list.html}} has since moved to an unknown location.

Despite the incremental progress that has been made, a large fraction of supernova data remains largely cloistered within the confines of individual supernova groups' web pages, if it is available at all, while a significant number of supernova enthusiasts either rely on David Bishop's Latest Supernova page, or Skywatch\footnote{\url{http://skywatch.co/skywatch}}, for metadata collected (and reported) by today's supernova surveys.  

Of the archives and catalogs in place, many also come with longstanding pitfalls and limitations for timely analysis. Primarily, there is little guarantee that data either donated to or sought after by such repositories will make it into public hands at the time of publication, or relatively soon thereafter. As a result, even estimating the total amount of public supernova data that is actually available is difficult, as it can only be bounded by the largest comprehensive catalogs that are available, e.g., WISeREP's collection of supernova spectra sans the dozens (if not 100s) of duplicate copies generated from including both rapid and final reductions of spectroscopic data. 

With the growing number of discoveries every year, as well as the millions expected during the era of the Large Synoptic Survey Telescope, there is a clear need for a complete data repository. This conglomerate of data would ideally span follow-up observations from both the most uniform supernovae (e.g., spectroscopically ``normal'' SN~Ia; \citealt{Graham15}) and the diverse subsets among thermonuclear, core-collapse, and more exotic events. For the benefit of reproducability of scientific results, there is also an impetus to make the data that was used to come to the conclusions presented in published literature publicly available.

Unfortunately academic journals and institutions have yet to enact a collection policy for published supernova data. Consequently, unless the data happens to be donated to a repository some years or even decades after the data were obtained (e.g., the day~$+$611 spectrum of SN~1994D; \citealt{Blondin12}), such observations have the potential to be lost indefinitely. In addition, published observations that have not been made public implies that: some theoretical models that have been deemed successful for select events are not also constrained by the wealth of other spectroscopically similar observations; numerous observations have yet to be fully explained by any model; and these missing, published datasets have typically only been analyzed once by a single group (or even single person), and to varying degrees of completeness.

In this work we present the \osc, which represents an ongoing community-driven project to collect and clean supernova metadata and observations spanning X-ray, ultraviolet, optical, infrared, and radio frequencies. At present, $\sim$14,000 events in the catalog include light curves with at least 5 photometric points, and in total the catalog contains $\sim$400,000 individual photometric detections. Within the catalog there are $\sim$5000 supernovae that include spectra, with $\sim$16,000 spectra in total.

In Section \ref{sec:principles} we overview the guiding principles for the \osc. In Section \ref{sec:capabilities} we outline the current capabilities of the catalog and describe how users can contribute their own data to the catalog. In Section \ref{sec:utility} we discuss present an incomplete list of ideas for potential applications of the catalog. In Section \ref{sec:future} we comment on our ongoing and future work, including an estimate of the quantity of data that the catalog is missing, and where we might find such data. In Section \ref{sec:summary} we summarize and provide final remarks.

\section{Guiding Principles}\label{sec:principles}
In creating this catalog, we aspire to meet a set of guiding principles that we believe form the basis for an objectively useful resource. While it is difficult to meet all these criteria simultaneously, we believe the general system we have selected is capable of at least pushing us towards the ultimate goal of making all public supernova data as readily accessible as possible.

The objects we include in the catalog are intended to be entirely {\it supernovae}, i.e. the complete destruction of a star by an explosive event that may or may not leave behind a compact remnant, and we actively remove objects that have been definitively identified as other transient types. One difference between our approach and some other supernova catalogs is that we augment the known supernovae with known supernova remnants \citep{Green:2014a,Maggi:2016a}, which are thought to be supernovae but (currently) with no known associated transient. As records continue to be studied and reveal historical observations \citep[see e.g.][for a recent example]{Neuhaeuser:2016a}, the prospect that these remnants will one day be associated with newly uncovered data remains an enticing possibility, and thus we believe that these remnants should be included amongst the observed supernova transients for completeness.

\subsection{Completeness and Persistence}
One of the primary objectives for the \osc is to have a complete collection of public data. For supernovae, this includes near-visible spectra and photometry (i.e. UV, optical, IR), as well as observations at X-ray and radio frequencies. Our goal is to include data not only from the latest supernovae, but also from the very oldest, such as the light curves and spectra of SN 1572\footnote{\url{https://sne.space/sne/SN1572A/}}, i.e. Tycho's supernova \citep{RuizLapuente:2004a,Krause:2008a}, such that the \osc becomes a reliable first destination for exploring the supernova dataset.

Equally important is that once the available data is included in the repository, it remains publicly available alongside all observations for a given event, even in the event that the creators of the catalog (the authors) do not actively maintain it anymore. This is facilitated by keeping as much of the data as possible on a freely-available service such as \github as the dataset can be cloned and forked by other users to provide data redundancy and permanence. Even if \github (or the authors) ceased to exist, our intention is to have cloned copies of the \osc on as many hard drives as possible, whereas currently the data associated with many historical supernovae have zero redundancy, i.e.~a single copy exists on a single hard drive.

\subsection{Community Driven}

Any group or individual can donate their data once published and/or made public to the \osc. To jump-start the catalog, we have collected supernova data from numerous sources on our volition, but as the number of events grows, such an approach will only be sustainable with the active participation of the community. 

To maximize community participation, we offer several different ways for users to donate data which are summarized on our contribute page\footnote{\url{https://sne.space/contribute/}}, with the simplest being a simple \texttt{Dropbox} file request through which users can upload folders of data. Small edits are possible directly through the main catalog interface by clicking an icon next to each event's entry (see Figure \ref{fig:maincatalog}), this takes the user directly to a page where data contributions can be made via a pull request. Finally, the user can submit their data in the specific \json format we have adopted. 

Our intention here is to minimize barriers for the data contributor as much as possible, the onus is upon the \osc to sanitize and agglomerate the donated data. The main factor that the mode of donation will affect is turn-around time, with high-quantity, low-effort data donations (e.g. photometry from dozens of supernovae) being added before low-quantity, high-effort donations (e.g. a single spectrum from a single event). We also respect that many will simply choose to continue to submit their data to other public repositories of supernova data, and we welcome greater connectivity with other services such that all the publicly-available data are accounted for.

\subsection{Accuracy and Accountability}

The \osc takes advantage of \git's distributed revision control and source code management, in addition to features of the freely-available \git hosting service \github. This enables collaborative corrections to be distributed for either incomplete data, or data in need of recalibration, observer frame corrections to spectra, photometric corrections, revisions of metadata, deletions of duplicate observations, and so forth. Anyone can either raise issues or suggest changes to any file for any project directory through \github's integrated issue tracking. This can be done either through the \github repository\footnote{\url{https://github.com/astrocatalogs/astrocats/issues}}, or by merging one's own local ``forked'' repository with the suggested changes. 

For every source of data we require a reference, which is preferably a published work (identified uniquely by an ADS bibcode) but is not required to be. Often times we draw from other online catalogs that themselves collected the data from another source, in such cases we explicitly trace the source chain back to the originating source of data, denoting the intermediary sources as ``secondary'' sources, a designation that also applies to the \osc.

\subsection{Accessibility}
A key factor for the success and impact of any catalog is access. In the supernova community rapid access is particularly useful, as important physical phenomena, such as shock breakouts or interactions with a nearby companion, are manifest in the early-time behavior of a supernova's evolution \citep{Falk77,Klein78,Kirshner87,Campana06,Soderberg08D,Schawinski08,Bloom12,Garnavich16}. Given that only a finite amount of glass is available to cover the full sky at any time, with many professional sites often being impacted by daylight, weather, or technical issues, the supernova community is often reliant upon amateur astronomers to fill the temporal and spatial gaps. For example, amateur astronomers were crucial in constraining the distance and light curve parameters of SN~2011fe (\citealt{Tammann11,Richmond12}, see also \citealt{Vinko12,Munari13,Pereira13}). 

Access is also important for public outreach, indeed many of the primary scientific results involving supernovae remain behind paywalls and are not available in a form the public can view. Even the simplest of questions the public might ask (how many supernovae have been discovered, for instance) remained difficult to answer prior to the creation of the \osc. To maximize access to the dataset produced by the \osc, we make it available for download at any time via the output repositories\footnote{\url{https://sne.space/download/}} within all individual supernova \json files are stored.

\subsection{Integration}
As the data products of the \osc are all stored on \git repositories, all of the supernova data collected can be downloaded locally, enabling analyses that are only limited by the user's available resources. The completeness of data allows anyone to filter and interrogate the latest available data without needing to first collect and plot it themselves, a time-saver that avoids having new accessors of a given dataset repeatedly perform the same data extraction procedure.

A recent example of such a data extraction prior to the existence of the \osc is \citet{Sasdelli16} where type-Ia supernovae were collected from a variety of sources and presumably homogenized; in principle this exercise only needs to be done once, and it is ideal for data accuracy reasons to perform this homogenization in the open where the process can be double-checked. In turn, the \osc gives supernova enthusiasts a unique opportunity to access all available data, query data statistics, and engage in an open dialog about the quality of that data.

\section{Catalog Capabilities}\label{sec:capabilities}

The \osc is built inside a broader project called \astrocats\footnote{\url{https://github.com/astrocatalogs/astrocats/}} which is being developed as a simple framework for constructing general astronomical catalogs like the \osc.  \astrocats is a python package providing a simple, nested structure of dictionary-like base-classes which mirror the eventual \json output files.  The core data-manipulation routines of the \osc use objects sub-classed from \astrocats, and extended with additional, supernovae-specific functionality (e.g.~accommodating standard supernova naming conventions, and distinguishing between photometry of the event and that of the host galaxy). 

\begin{figure}[t!]
\centering
\begin{tikzpicture}
\begin{scope}
[node distance=0.5cm and 0cm]
\node (observers) [user] {\footnotesize Observers};
\node (externallive) [web, below= of observers] {\footnotesize Internet sources};
\node (externalstatic) [repo, left= of externallive, xshift=-1cm] {\footnotesize Static};
\node (internal) [repo, right= of externallive, xshift=1cm] {\footnotesize Internal};
\node (import) [process, below= of externallive] {\footnotesize AstroCats: {\tt import}};
\node (eventjson) [file, below= of import] {\footnotesize Event \json files};
\node (makecatalog) [process, below= of eventjson] {\footnotesize AstroCats: {\tt webcat}};
\node (catalogrepo) [repo, left= of makecatalog, xshift=-1cm] {\footnotesize Catalog};
\node (eventrepos) [repo, right= of eventjson, xshift=1cm] {\footnotesize SNe};
\node (catalogjson) [file, below= of makecatalog] {\footnotesize catalog.json};
\node (osc) [web, below= of catalogjson] {\footnotesize \bf Open \mbox{Supernova} Catalog};
\node (eventpages) [web, right= of osc, xshift=0.5cm] {\footnotesize Individual SNe pages};
\node (webrepo) [repo, left= of osc, xshift=-1cm] {\footnotesize Website};
\node (scientists) [user, below= of osc] {\footnotesize Scientists};
\draw [arrow] (observers) -| (externalstatic);
\draw [arrow] (observers) -- (externallive);
\draw [arrow] (observers) -| (internal);
\draw [arrow] (externalstatic) |- ([yshift=0.2 cm]import.west);
\draw [arrow] (externallive) -- (import);
\draw [arrow] (internal) |- (import);
\draw [arrow] (import) -- (eventjson);
\draw [arrow] (eventjson) -- (makecatalog);
\draw [arrow] (makecatalog) -- (catalogjson);
\draw [arrow] (eventjson) -- (eventrepos);
\draw [arrow] (catalogrepo) -- (makecatalog);
\draw [arrow] (catalogrepo) |- ([yshift=-0.2 cm]import.west);
\draw [arrow] (catalogjson) -| (catalogrepo);
\draw [arrow] (makecatalog) -| (eventpages);
\draw [arrow] (catalogjson) -- (osc);
\draw [arrow] (webrepo) -- (osc);
\draw [arrow,dashed] (osc) -- (eventpages);
\draw [arrow] (eventpages) |- ([xshift=-0.1cm, yshift=0.2 cm]scientists.east);
\draw [arrow] (osc) -- (scientists);
\draw [arrow] (eventrepos.east) |- ([xshift=-0.1cm, yshift=-0.2 cm]scientists.east);
\end{scope}
\end{tikzpicture}
\caption{Flow chart showing how sources of data are combined and used to generate the catalog webpage and its datafiles, where the yellow ellipses are human users, the green diamonds represent \git repositories, the blue trapeziums represent internet web pages, the orange rectangles are \python methods, and the red rounded rectangles are \json files. The solid arrows represent direct input of data in their indicated direction, whereas the dashed arrows represent hyperlinks.}
\label{fig:flowchart}
\end{figure}
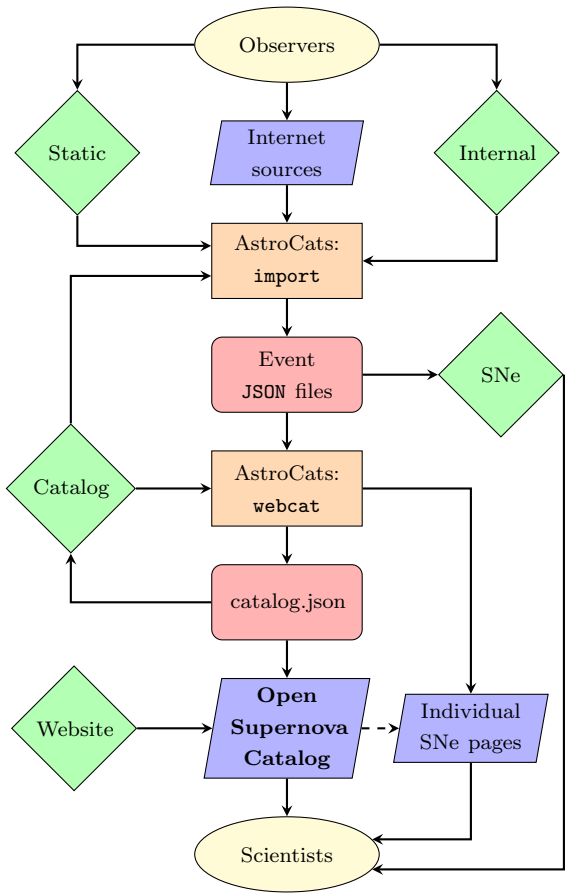

In this section, we briefly describe the features of \astrocats as it applies to the \osc and leave details of the general framework and implementation to a future paper.  There are two critical methods that underlie the \osc's functionality: `{\tt import}', which both agglomerates data from a variety of online and published sources, and generates individual \json files for each supernova; and `{\tt webcat}', which consolidates all supernovae into a single, \json catalog-file and then generates the web documents by which that data will be displayed. In Figure~\ref{fig:flowchart} we show how these methods (represented by orange rectangles) pull from the available sources of data, produce outputs, and interact with one another to make the catalog possible.

\subsection{Human-Readable JSON Files}
All supernovae within the \osc are stored in individual \json files which bear the name of each event (e.g.~SN~1987A is contained within a \json file simply entitled `SN1987A.json'), and the catalog file that collects all supernova metadata is also stored in the \json format. \json (indicated by the red rounded rectangles in Figure~\ref{fig:flowchart}) is a serialized, hierarchical format with minimal complexity, and is typically encoded as human-readable ASCII, although encoding binary data is possible within the format. Because of its simplicity, there exist hundreds of \json interpreters in dozens of programming languages\footnote{\url{http://json.org}}. The hierarchical nature of \json means that fields can be accompanied by sibling or child fields that can provide additional information on a datum, such as the source (e.g.~citation) of that datum or the error value associated with it.

Unlike the two-dimensional ASCII tables that supernova data is often presented in, data presented in hierarchical formats like \json can be sparse; i.e.~not all values need to share the same set of fields. For instance, the instrument on a telescope used to collect data may be known for one observation but not another. With a sparse format the observation without this information can simply omit the {\tt instrument} field. Often times the omission of data in a standard ASCII table is accomplished by assigning a ``placeholder'' value, as an example the redshift value for a supernova with an unknown redshift may be filled with an unreasonably large value such as 999. Because these placeholders are arbitrary and vary greatly from source to source, it is preferable to omit them as they can potentially contaminate datasets for those who are not intimately familiar with the placeholder convention that the producer of that data has adopted.

\json's advantage over pure binary formats is its readability; i.e., a \json file can be easily checked for errors by eye, whereas a binary-encoded file (e.g. {\tt FITS}) requires translation by an interpreter before it can be inspected for errors. When disk space is at a premium, storing data as binary can be significantly more efficient than ASCII, but much of this benefit is mitigated by simply compressing the ASCII files, which most web servers perform before delivering data to a visitor anyway. As an example, our largest \json file presently belongs to SN~2013dy, the uncompressed \json file for this event is 144MB in size, whereas the version compressed with {\tt gzip} is only 22MB, which is almost as small as the data would be in pure binary format ($\sim$ 10MB), and requires only a fraction of a second to read from increasingly ubiquitous solid state drives.

A similar format to \json is \xml, which is more powerful but also more verbose than \json, and is currently the markup of choice for the VOTable\footnote{\url{http://www.ivoa.net/documents/VOTable/}} and VOEvent \citep{Seaman:2011a} standards. The disadvantages of \xml are that it uses somewhat more space than \json as both opening and closing tags are required for every field, and that the format is intrinsically more complicated than \json and is thus more difficult for humans to read, which in the opinion of the authors makes it less accessible to those not familiar with its syntax. That being said, it is trivial once the data has been standardized into a serialized format like \json to export it to another like \xml, and as we describe in Section \ref{sec:future} we plan to add VOTable outputs for each event as complements to our \json database.

\subsection{Client-Side Tabular Interface}
For catalogs of data intended for web delivery, a choice needs to be made whether the catalog will be stored client-side or server-side. Server-side catalogs exchange a minimum of data with the client and are only limited by the processing capacity of the server, and can enable interaction between the user and catalogs that may contain billions of entries. Client-side catalogs are limited by the bandwidth of the server and the computing power of the client's computer, and are practically limited to smaller catalog sizes that the client's computer can handle. The problem of storing the whole of astronomical data, especially in the upcoming era of wide-field surveys, is an immense computational challenge, and the machines that serve such data will need to consist of many thousands of CPUs and many petabytes of disk space, which favors the server-side model.

However, for a given astronomical sub-field, the total number of objects is typically much more manageable, and even for an extremely mature field, e.g. observations of supernovae, the total number of objects rarely exceeds the tens of thousands. For such a sample size, the total metadata content in uncompressed form is typically only tens of megabytes (MB), and only a few MB after data compression (e.g. {\tt gzip}). While MB of data posed issues prior to the ubiquity of broadband internet, such a catalog can be delivered and rendered by the client in seconds on modern-day networks, and with optimization, catalogs of hundreds of thousands of objects can still yield better performance and usability than equivalent server-side solutions.

Once the data has been delivered to the client, data access is only limited by the processing power and memory speed of the client, this results in a significantly smoother user experience than server-side solutions which can be excruciatingly slow when overloaded, as is typical of many under-funded and under-supported web resources that astronomers and astrophysicists rely upon. Searches and filtering of the data can be orders of magnitude faster when performed client-side, a difference that can result in greatly enhanced productivity, especially in the early phases of a project when a scientist is forming and testing their ideas.

\begin{figure*}
\centering
\includegraphics[width=1.\linewidth]{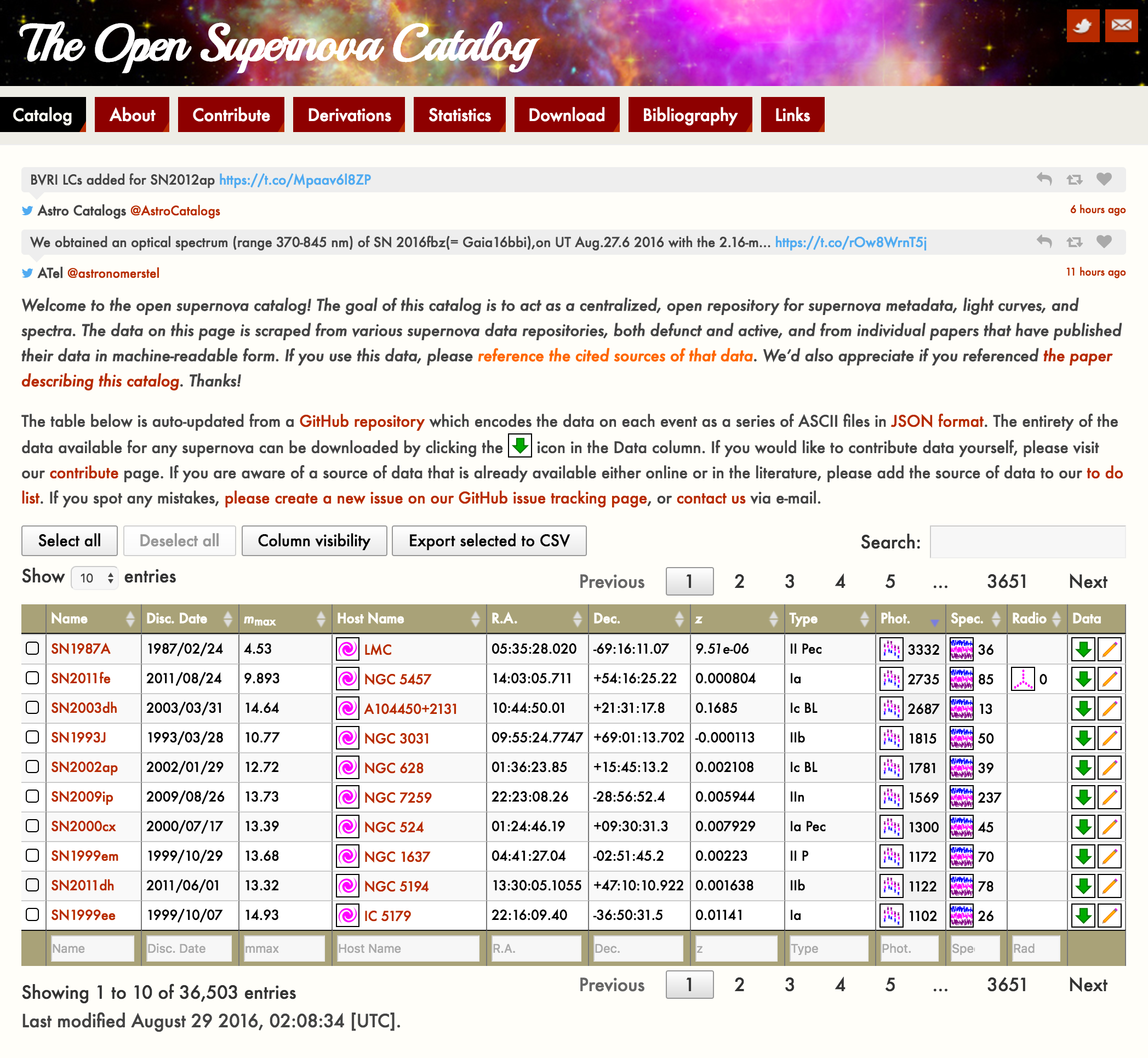}
\caption{Screenshot of main interface as it appears on the homepage of the \osc, \url{https://sne.space}. The front-facing table is intended to be a user-friendly, rapidly-searchable database of supernovae, with instant access to the full dataset associated with each event via the download button (the down-facing arrow) in the rightmost column. Each event can be edited directly by clicking the ``pencil'' icon in the rightmost column, which takes the user to a \github edit page for that particular event. The search bar in the upper right performs simple searches of all columns, whereas the fields at the bottom of the table enable searching of each individual column. Counts of the amount of photometric detections and available spectra are displayed in their own columns. Additional columns which are not shown by default (such as event aliases, luminosity distance, etc.) are accessible via the ``column visibility'' button. The webpage has been optimized for both desktop and mobile browsers and loads the full metadata catalog in its entirety in seconds on a broadband connection.}
\label{fig:maincatalog}
\end{figure*}

For the \osc, our tabular web interface is powered by the \dt software package written for {\tt jQuery}, a widely-used JavaScript library for creating online interactive tables (Figure \ref{fig:maincatalog}). \dt can be run in both client- and server-side modes, and we choose client-side for the reasons stated above. While the client-side approach should scale to hundreds of thousands events without issue, the full supernova population may number in the millions in a decade, at which point a complete conversion to a server-side model, or even a hybrid client/server-side model, would be possible. If such a conversion is performed, {\it the same level of accessibility to the catalog and its data must be retained}.

\subsection{Version Control with \git}
Whereas \dt is the face of the \osc, its heart is \git, a standard version control software package. The entirety of the catalog, including its interface software, input data, the individual \json event files, the primary catalog \json file, and the scripts written to generate the catalog and event files, are stored as \git repositories on \github (green diamonds in Figure~\ref{fig:flowchart}). This offers an unprecedented level of detail and accountability for the data presented on the \osc, enabling precise determinations of when its contents changed, how they were changed, and who was responsible for the changes. This is in stark contrast to closed catalogs which at best may track changes internally but offer no mechanism for external scrutiny.

In constructing the \osc we have found dozens of errors in both published and unpublished sources, and continue to find errors on a regular basis (see Section~\ref{sec:metadata}). Once these errors are found they are documented as issues on \github and eventually corrected, but in principle these errors can easily propagate to future works if left unchecked. When the data is stored in a version control environment such as \git, the best practice for a paper that utilizes aggregated or generated data is to use the commit hash associated with the version of the data they used. And if errors are corrected after publication that may affect a publication's results, the commit history of the data can be inspected to determine what adjustments are required such that the propagation of the error is minimized.

\subsection{Adding Data and its Sources}\label{sec:addingdata}
Two important aspects of collecting data is having a reasonable way of reconciling different data values from different sources, and ensuring that the genesis of the data that appears in the final product is not lost in the collection process. For any data quantity to be added to the \osc, the quantity {\it must} have a citation before the import script will allow the value to be added. If the value agrees with the value reported by another source, the system links that value to both sources.

In practice this is done by maintaining a {\tt sources} array in the individual event \json files which details all sources of data used to generate the file, and then adding a {\tt source} property to every data quantity which refers back to the event's source list. A full schema description is available within the primary \osc repository\footnote{\url{https://github.com/astrocatalogs/supernovae/blob/master/SCHEMA.md}}, which should always be checked for the latest schema. Below we provide an example of what the schema looks like in the \json file for SN2011fe at the time of this paper's publication, ellipses are shown where data has been omitted for brevity:

\begin{Verbatim}[obeytabs,tabsize=2,fontsize=\footnotesize]
...
"schema":".../SCHEMA.md",
"name":"SN2011fe",
"sources":[
	...
	{
		"name":"Maguire et al. (2014)",
		"url":"...",
		"bibcode":"2014MNRAS.444.3258M",
		"alias":"3"
	},
	...
	{
		"name":"Asiago Supernova Catalogue",
		"url":"...",
		"alias":"9",
		"secondary":true
	},
	{
		"name":"ATel 3581",
		"url":"...",
		"alias":"10"
	},
	{
		"name":"Latest Supernovae",
		"url":"...",
		"alias":"11",
		"secondary":true
	},
	...
],
...
"ra":[
	{
		"value":"14:03:05.76",
		"source":"3",
		"unit":"hours"
	},
	{
		"value":"14:03:05.81",
		"source":"9",
		"unit":"hours"
	},
	{
		"value":"14:03:05.80",
		"source":"10,11",
		"unit":"hours"
	}
],
...
"photometry":[
	{
		"time":"55799.0045",
		"u_time":"MJD",
		"band":"W2",
		"system":"Vega",
		"magnitude":"17.579",
		"e_magnitude":"0.152",
		"telescope":"Swift",
		"instrument":"UVOT",
		"source":"8"
	},
	...
	{
		"time":"55798.00",
		"u_time":"MJD",
		"frequency":"8.5",
		"u_frequency":"GHz",
		"fluxdensity":"0.0",
		"e_fluxdensity":"25.0",
		"u_fluxdensity":"µJy",
		"instrument":"VLA",
		"source":"2"
	},
	...
	{
		"time":["55800.443", "55801.042"],
		"u_time":"MJD",
		"energy":["0.5", "8."],
		"u_energy":"keV",
		"flux":"7.5e-16",
		"unabsorbedflux":"7.7e-16",
		"u_flux":"ergs/s/cm^2",
		"photonindex":"2.",
		"counts":"1.1e-4",
		"upperlimit":true,
		"instrument":"Chandra",
		"nhmw":"1.8e20",
		"source":"4"
	},
	...
],
...
"spectra":[
	{
		"instrument":"LT - FRODOspec",
		"u_time":"MJD",
		"time":"55797.0",
		"filename":"11kly_20110824_LT_v1.ascii",
		"waveunit":"Angstrom",
		"fluxunit":"erg/s/cm^2/Angstrom",
		"data":[
			["3900.399902", "9.784687E-15"],
			...
		]
        "source":"10,23"
	},
	...
]
...
\end{Verbatim}

In the above example, three different values for the right ascension of SN2011fe have been reported by four different sources. Two of the sources are tagged with a {\tt secondary} Boolean value to indicate that they themselves are not the original sources of this data, but collected the data from other sources. In this way, the full source chain by which a value was acquired can be traced back to its original source, and the user of the catalog can decide which value they would like to utilize when performing data analysis. As explained in Section \ref{sec:cleanup}, not all data is added from all sources, we perform some basic quality control before a piece of information is appended to a supernova. A complete table of all published sources we draw from is available via a bibliography table that is constructed from the catalog\footnote{\url{https://sne.space/bibliography/}} (4000+ published sources as of May 2, 2016).

Photometry and spectra have a separate set of tags unique to each of them that provides all known data regarding a particular collected observation, which helps identify similar datasets that may refer to the same set of observations. For photometry, an observation can be reported in terms of a {\tt magnitude} in a particular {\tt band}, {\tt flux}, or {\tt fluxdensity} depending on whether the observation is IR/optical/UV, radio, or X-ray; one of each case is presented in the example above. As photometric errors are closer to Gaussian in flux-space, photometric observations also support the {\tt flux} and {\tt count} tags which are used when the original source provides these values; unfortunately, the majority of historical supernova photometry has been presented in magnitudes. Spectra are tagged with the {\tt filename} they originated from, as spectral data is usually delivered in the form of a single {\tt FITS} or {\tt ASCII} file per spectrum, this provides another consistency check for the origin of a given spectrum in addition to the {\tt sources} tag. Both photometry and spectra are tagged with {\tt survey}, {\tt observatory}, {\tt observer}, {\tt reducer}, {\tt telescope}, and {\tt instrument} tags if these are known.

\subsection{Individual Supernova pages}

\begin{figure}[t!]
\centering
\includegraphics[width=\columnwidth, trim= 0mm 0mm 0mm 0mm]{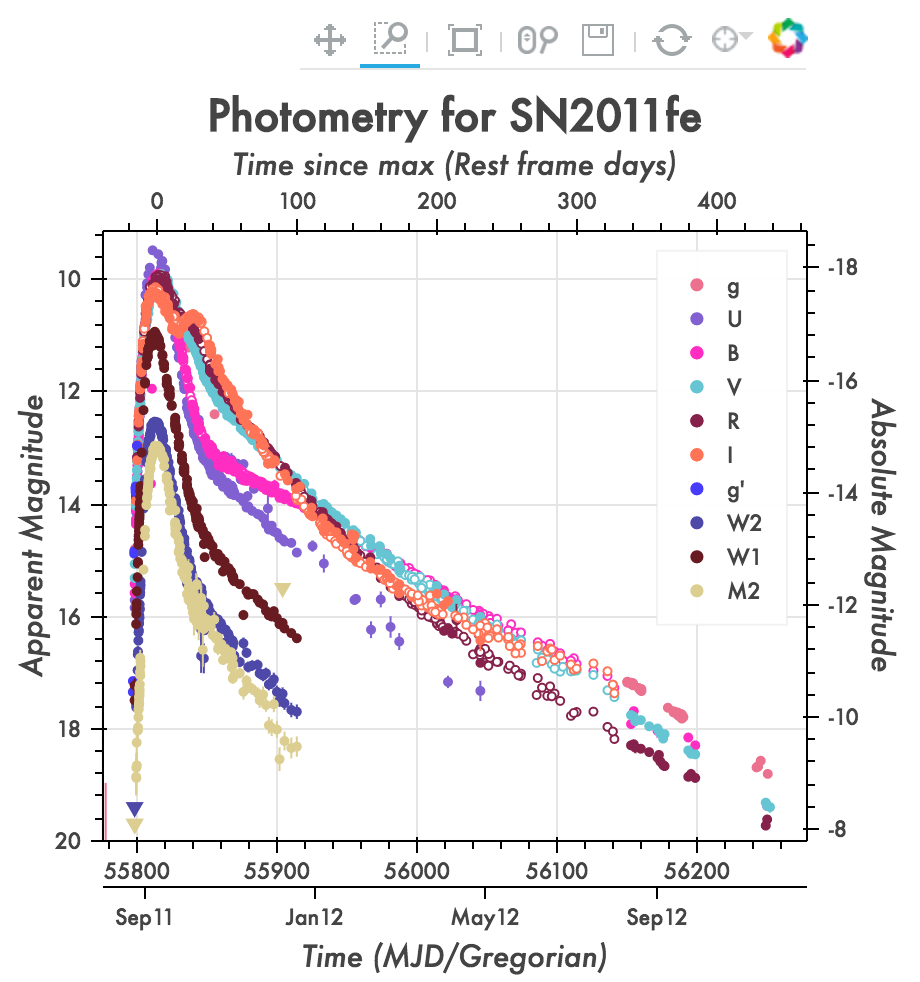}
\caption{Screenshot from interactive plotting of photometry for SN~2011fe, available at \mbox{\url{https://sne.space/sne/SN2011fe}}.}
\label{Fig:11fe_phot}
\end{figure}

\begin{figure}[t!]
\centering
\includegraphics[width=0.9\columnwidth, trim= 0mm 0mm 0mm 0mm]{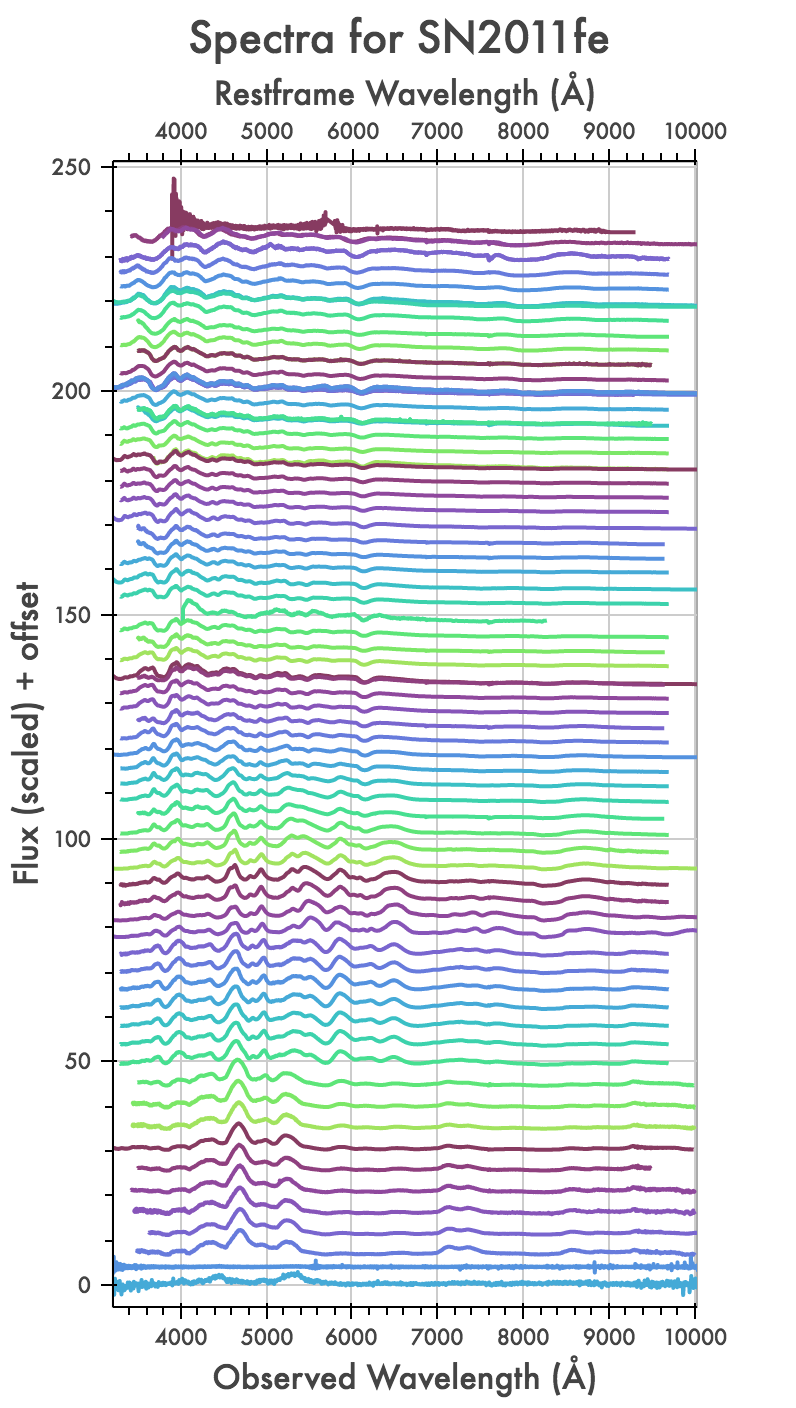}
\caption{Screenshot from interactive plotting of time-series spectra for SN~2011fe, available at \mbox{\url{https://sne.space/sne/SN2011fe}}.}
\label{Fig:11fe_specs}
\end{figure}

For supernovae where observations have been made available, these data and the related metadata are displayed on individual event pages. Shown in Figures~\ref{Fig:11fe_phot}~and~\ref{Fig:11fe_specs} are examples of the graphical interfaces used by the \osc, in this instance for most of the publicly available observations of the well-observed SN~2011fe \citep{Nugent11, Richmond12, Parrent12, Matheson12, Vinko12, Maguire12, Pereira13, Mazzali14, Maguire14, Mazzali15nebular11fe, Taubenberger15, Graham15, Smitka16}. 

\begin{figure*}[t]
\centering
\includegraphics[width=0.4\linewidth, trim= 0mm 0mm 0mm 0mm]{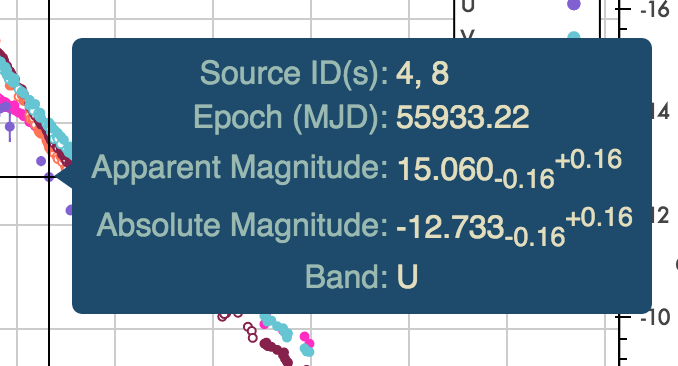}~~
\includegraphics[width=0.33\linewidth, trim= 0mm 0mm 0mm 0mm]{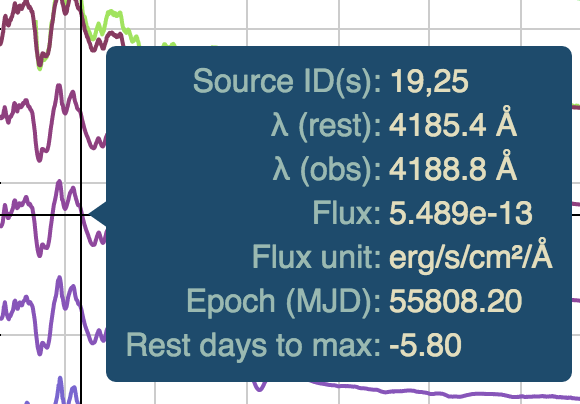}
\caption{Screenshots of additional information available when hovering over individual photometric observations (left panel) and individual spectra (right panel) in the interactive plotting for SN~2011fe, available at \mbox{\url{https://sne.space/sne/SN2011fe}}.}
\label{Fig:hoverexample}
\end{figure*}

Photometric points are presented in terms of both apparent and absolute magnitudes along with the observer-frame time relative to MJD and the Gregorian calendar. Along with the magnitudes in a given band filter, additional bits of metadata such as the observatory, telescope, and instrument used to collect the data are also included, as well the photometric system the magnitudes are reported in. Time-series spectra are plotted sequentially within a single interactive window, which enables anyone to inspect select spectral features and their respective evolutions prior to and after maximum light (see, e.g., \citealt{Black16}). Within each graphical display, a user can mouse-over individual photometric and spectroscopic data and recover select pieces of relevant information, e.g., the number of days since maximum light for a given spectrum (see Figure~\ref{Fig:hoverexample}). Spectra are presented in both the observer and rest frames 1 except in certain cases where the frame is not known.

For each radio observation we report the start and end time (in MJD, observer frame), the central frequency of observation (in GHz, observer frame), the measured flux density at that frequency at the transient position (in micro Jy) and the $1\sigma$ rms (in micro Jy). X-ray observations include both the start and stop time (in MJD, observer frame), the energy band utilized (in keV, observer frame); the measured count-rate; the observed flux (i.e. absorbed, both by the Galaxy and intrinsically, units of $\rm{erg\,cm^{-2}\,s^{-1}}$) and related uncertainty ($1\sigma$ confidence level, unless stated otherwise); the Galactic neutral hydrogen column density used in the fit (units of $\rm{cm^{-2}}$); the best-fitting intrinsic neutral hydrogen column density at the redshift of the source (units of $\rm{cm^{-2}}$); the unabsorbed flux (where both the Galactic and the intrinsic absorption have been accounted for) and associated uncertainty ($1\sigma$ confidence level, unless stated otherwise); and information about the spacecraft/instrument and the reference to the source of data. The spectral model used to fit the data and derive the fluxes above is indicated with an acronym (e.g.~PL stands for power-law). The best-fitting values of the other parameters of the fit (and their uncertainties) are also listed. Upper limits are provided at the $3\sigma$ confidence level, unless otherwise noted.

\subsection{Contributing Data}\label{sec:contributing}
Because we have already collected data from most existing public repositories, much of the catalog's future growth is now dependent upon individual donations. At the moment we offer four different ways for users to contribute their data, with the main intention being a minimization of the amount of effort on the part of both the submitters and receivers of that data. As we describe in \ref{sec:future}, the particular means of data contribution we summarize here may evolve to address any community concerns and to make data contribution even simpler, so users of the \osc should always check our data contribution page\footnote{\url{htps://sne.space/contribute/}} for the latest instructions on how to contribute data.

\begin{enumerate}
\item For photometry and metadata, we highly recommend that users submit their data to the VizieR service and then inform us when it is available, either via our to do list or via e-mail. Once in VizieR, it is trivial to pull the data into the \osc, often times with only a couple lines of \python. This way, your data is not just available via the \osc, but also to the broader community who may not know the \osc exists.

\item Small contributions of data can be provided manually by clicking the edit link (the pencil icon, see Figure \ref{fig:maincatalog}) next to each entry on the main catalog page. This will take you directly to a \github edit page where you can add data directly in the \json format. When this edit is saved, \github will automatically submit a pull request to our ``internal'' GitHub repository\footnote{\url{https://github.com/astrocatalogs/sne-internal}}. When data is contributed in this way, the submitted \json file should conform to our format specification (see Section~\ref{sec:addingdata}), with any formatting issues being mediated via the pull request system.

\item The simplest way for a user of the \osc to submit data is to upload the data to a Dropbox file request we provide on the contribution page. We have no specific format requirements for such donations, but use of standard file formats will expedite the addition of data to the catalog and is highly recommended. All uploads should include a {\tt README} file that describes the data, with the {\tt README} containing the basic contact info of the uploader. The data should also include at least one reference, preferably with an associated ADS bibcode, for source attribution. The data itself should preferably be human-readable ASCII (\json, {\tt CSV}, {\tt XML}, etc.), but if not, a \python script should be included that will read the data into memory.

\item If you are planning bulk contributions to the catalog but do not want to convert your data to the \json format, the preferred way is to submit a pull request to one of our external data repositories, and optionally editing the import script itself to parse that data. Editing the import script itself might be a preferable solution if you are in charge of a data source that updates on a regular basis, as you will be able to adapt the importation to any changes that are made to the source. Because of \github repository size limits, the import script\footnote{\url{https://github.com/astrocatalogs/sne/blob/master/scripts/import.py}} and the external data\footnote{\url{https://github.com/astrocatalogs/sne-external}}$^{,}$\footnote{\url{https://github.com/astrocatalogs/sne-external-radio}}$^{,}$\footnote{\url{https://github.com/astrocatalogs/sne-external-xray}}$^{,}$\footnote{\url{https://github.com/astrocatalogs/sne-external-WISEREP}}$^{,}$\footnote{\url{https://github.com/astrocatalogs/sne-external-spectra}} is split into a few different repositories.

\end{enumerate}

\section{Utility of a Comprehensive Catalog}\label{sec:utility}

A comprehensive catalog foremostly enables finding systematic errors, e.g., 
identifying the true type of an event that may have been misclassified;
understanding the current completeness of the data and optimizing 
data collection of future surveys; evaluating data 
quality of current samples, e.g., accessing spectroscopic coverage in terms 
wavelength and days since maximum light in order to refine 
future community-driven follow-up campaigns; and 
determining the mean number of papers per light curve and/or per cumulative set of 
spectra, i.e. how much effort was needed to generate a given set of data. 
Additionally, a comprehensive catalog reduces unnecessary overhead for studies 
with selection criteria, e.g., querying those events with available 
multi-wavelength observations in the X-ray and/or radio in order to understand 
the environments around supernovae.

Quick science is another outcome of a comprehensive catalog. Examples include: 
estimating the number of supernovae of a given type discovered and/or 
observed per year, and its dependence on redshift (cf. \citealt{Maoz12}); 
determining the frequency of supernova type by galaxy type, and its 
relation to galactocentric distances assuming a volume limited 
sample \citep{WLi11b,WangX13,Graur:2016b,Graur:2016a}; identifying noteworthy supernova 
factories, e.g., NGC 2770 \citep{Hurst99,Soderberg08D}; and using the available 
photometry to estimate a (pseudo-) bolometric light 
curve per event, where appropriate. 

A comprehensive catalog can also lead to 
the creation of more complex,
real-time diagnostics of the latest data. This can generally be anything from 
finding commonalities between seemingly different supernova types, to 
generating mean spectra and exploring spectroscopic diversity 
\citep{Benetti05,Branch05,Branch06,Branch07b,Branch08,Branch09,WangX09Subtype,Blondin12,Folatelli13,Maguire14,Sasdelli15PCA,Sasdelli16,Kromer16}.

\section{Ongoing Work and Future Features}\label{sec:future}
While we believe that the \osc already addresses many of our guiding principles (Section \ref{sec:principles}), there are a number of improvements that can be made to further meet those goals. In addition to the example uses highlighted in Section \ref{sec:utility}, there are certainly many questions on the nature of supernovae that have yet to be asked simply because we are only just now collecting the data into a single location. While we believe that the \osc in its present form should be useful to scientists, much work remains to be done to improve upon the presentation and quality of the data we have collected.

\subsection{Clean-up and Standardization}\label{sec:cleanup}

A catalog can choose to engage in wholesale collection of all available data without vetting that data for quality, duplicity, and obvious errors, and for the \osc we prefer to collect all data even if the data may have unresolved issues. However, one of the benefits of collecting all available data in one place is that aberrant values can be directly compared and used to establish a consensus. In the following sub-sections we describe our strategies for homogenizing, cleaning, and presenting data that originates from a wide variety of heterogeneous sources.

\subsubsection{Metadata}\label{sec:metadata}

Before adding values from a source to the catalog, we perform some basic quality control checks which ensures that the final \json file only includes the most accurate data, although we err on the side of inclusion when the relative quality of two differing data values is ambiguous. For any textual data (such as host galaxy names or supernova types), quantities are run through a synonym list appropriate to that data type and are always added unless the datum is known to be in error (i.e. a typo). For numerical data, values with attached error bars are retained over values without error bars, and for data without error bars, values with the most significant digits are preferentially retained. For dates, we retain values with the most chronological information, for instance a discovery date of 2014 would be dropped in favor of 2014/06/03 if this date is provided in another source.

Name resolution, i.e. when a supernova has more than one name in two or more surveys, is also done by combining alias information from all sources that provide such data. Events are merged such that supernovae all appear under the same name, with the IAU {\tt SNYYYYxxx} format being preferred, if it is available. If an event lacks an IAU name, the discovering survey name is preferred. As determining whether two events are the same or not often requires directly inspection of imaging, we offer a ``duplicate finder'' page\footnote{\url{https://sne.space/find-duplicates}} on the \osc so that observers can suggest to us which events should or should not be listed under multiple names.

Frequently two or more conflicting values will appear in different sources for the same supernova. Finding and resolving these conflicts is paramount to maintaining a high-quality dataset. Because often times the conflicts cannot be resolved without referencing the original images or the collected spectra (both of which may or may not be public), we provide another community tool, a ``conflict finder'' page\footnote{\url{https://sne.space/find-conflicts}}, to mark values as being erroneous so that they can be ignored on subsequent imports.

\subsubsection{Light curves}

Supernova photometry suffers from a few pathological issues that are difficult to resolve without amassing a large collection of data. Among these issues are: not knowing which photometric system magnitudes are reported in; whether the photometry has been S- and/or K-corrected or not; how and if the host galaxy light was subtracted; the definitions of the photometric filters employed; using fluxes and flux errors rather than magnitudes; and the lack of error bars. These details are often elucidated in the source material, but the inclusion of such details is haphazard and of varying verbosity. Our principle goal is to collect photometry in its presented form and to label that data with as much metadata as possible, as we believe the best practice is to let the users of that data apply whatever corrections they believe to be appropriate for their use case.

For future data collection, our stated preference is for the raw light curve data to be accompanied by color corrections that enable the ideal scenario of displaying each supernova unextincted at $z = 0$, but the data should always be delivered in its raw form with the corrections being provided as separate entries. While the \osc currently displays photometric data as submitted, the data that is displayed on an individual supernova webpage in the future will take advantage of any provided corrections such that different supernovae (or different datasets collected on the same supernova) can be compared by eye more directly. Such clean-up will only be performed if we are confident that the source of data did not perform any corrections themselves, and will only affect the visual appearance of the light curves on the individual supernova pages.

\subsubsection{Spectra}

Two longstanding issues with the quality of optical spectra have been the application of ambiguous redshift corrections and the removal of telluric features. Subtracting telluric features from supernova spectra is important for taking measurements of certain features, e.g., blends of \ion{O}{1}~$\lambda$7773 and \ion{Mg}{2}~$\lambda$7869, 7890. However, in the event that shared data has been ambiguously corrected for redshift, and undocumented as such, telluric features serve a useful purpose when resetting the spectrum to a proper observer frame. 

This issue was first noticed by the administrators of SuSpect during the 2000s, where some of the spectra were successfully reset to a proper observer frame in subsequent works. However, these data were later collected by WISeREP, and no corrections to those affected data have been made since at Suspect, nor at WISeREP, which has left significant overhead for anyone seeking to utilize some of these publicly available datasets. This issue of misaligned spectra has even trickled to more recent works, e.g. four misaligned spectra in Fig.~1 of the preprint (v1) of \citet{Sasdelli16}.

In the \osc we have taken the initiative to reset those affected data files once and for all, and with the help of telluric features assuming they have not already been removed. A number of spectra were also found to have bad wavelength solutions (10s of spectra); we have removed these from the catalog and have added them to a list of events for which the data will need to be either reacquired, or digitized from the original manuscripts.\footnote{See \url{https://github.com/astrocatalogs/supernovae/blob/master/TODO.md}}

For literature data that have already been corrected for redshift, e.g. CSP spectra and UCB files with a {\tt noz} label (i.e. ``no z remaining''), we have ensured that these data are not twice-corrected in the catalog. However, to avoid these potentially fatal impacts on supernova research, we ask that spectra donated to the \osc (or our external sources such as WISeREP) are left uncorrected for redshift.

Additionally we aim to clean the spectral data and maintain accurate metadata. This work includes: removing telluric features; rescaling spectra to published flux values, and/or photometry, as many spectra have been released without scaling to the photometry \citep{Silverman12maxlight,Modjaz14}; providing internally consistent dates of maximum light so that spectra can be compared by an appropriate phase; and correcting records of maximum absolute magnitudes. 

All corrections applied by the \osc to spectra are in the form of added metadata tags, e.g. an {\tt exclude} tag can be added to an existing spectrum to exclude ranges of wavelengths with noisy data. As with the photometry, these corrections do not involve manipulation of the original data files, whenever possible.

Because one of the goals of the \osc is to organize a complete sample of supernova spectra, we aim to remove duplicate data found on through external resources such as WISeREP from our time-series collection of spectra. These are typically the same spectrum for two different modes of reduction (rapid versus final). As WISeREP and the Transient Name Server will still continue to be a great source of raw spectral data, we will continue to collect the spectra posted there into the catalog in perpetuity, applying the corrections above in the form of attached metadata. Finally we will look to obtain unavailable published optical spectra and light curves that are known to be missing from the external sources we draw from (see Table~\ref{tab:missing} for an incomplete list). 

\subsection{Missing Data}\label{sec:missing}

\begin{figure*}[t]
\centering
\includegraphics[height=0.7\columnwidth]{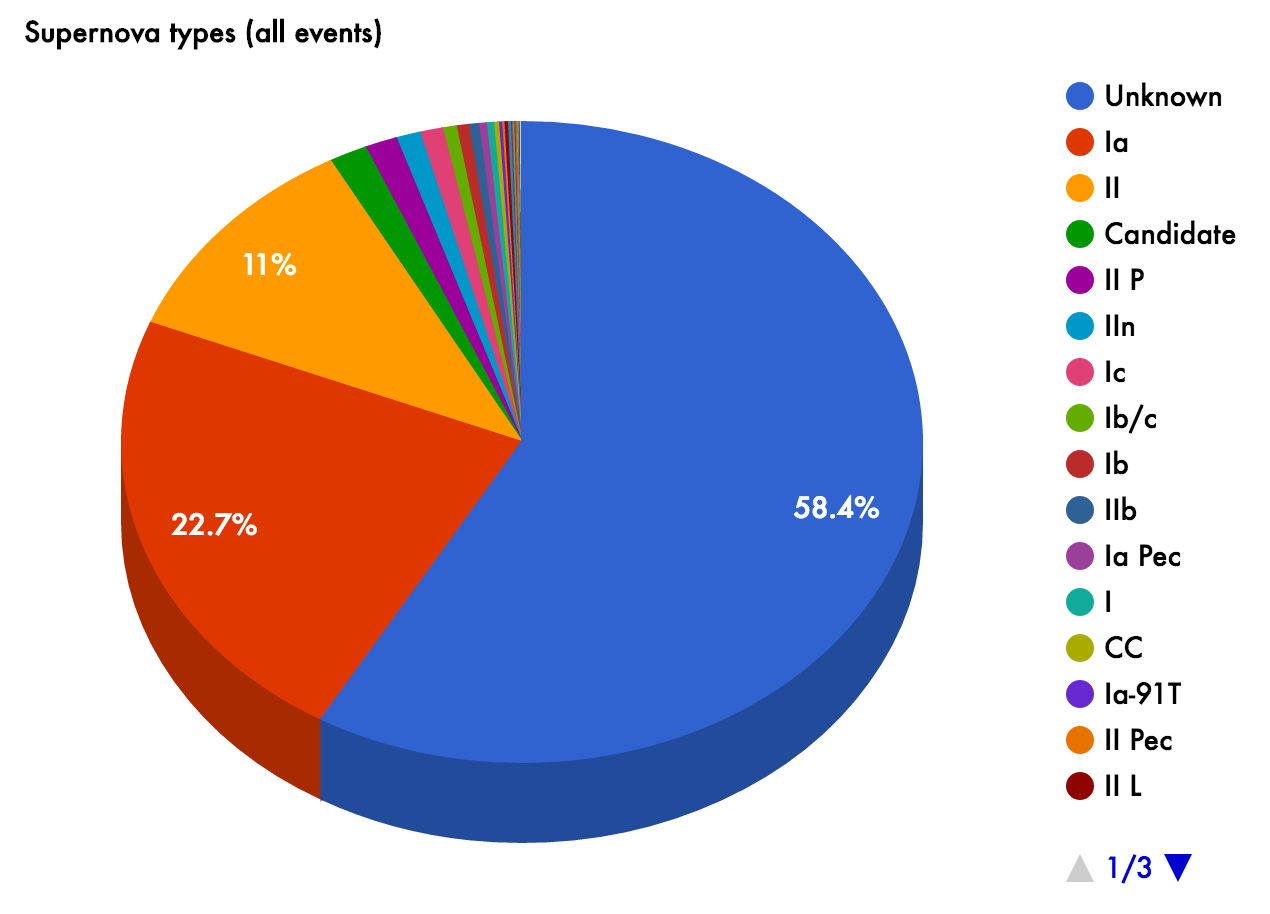}~~~~\includegraphics[height=0.7\columnwidth]{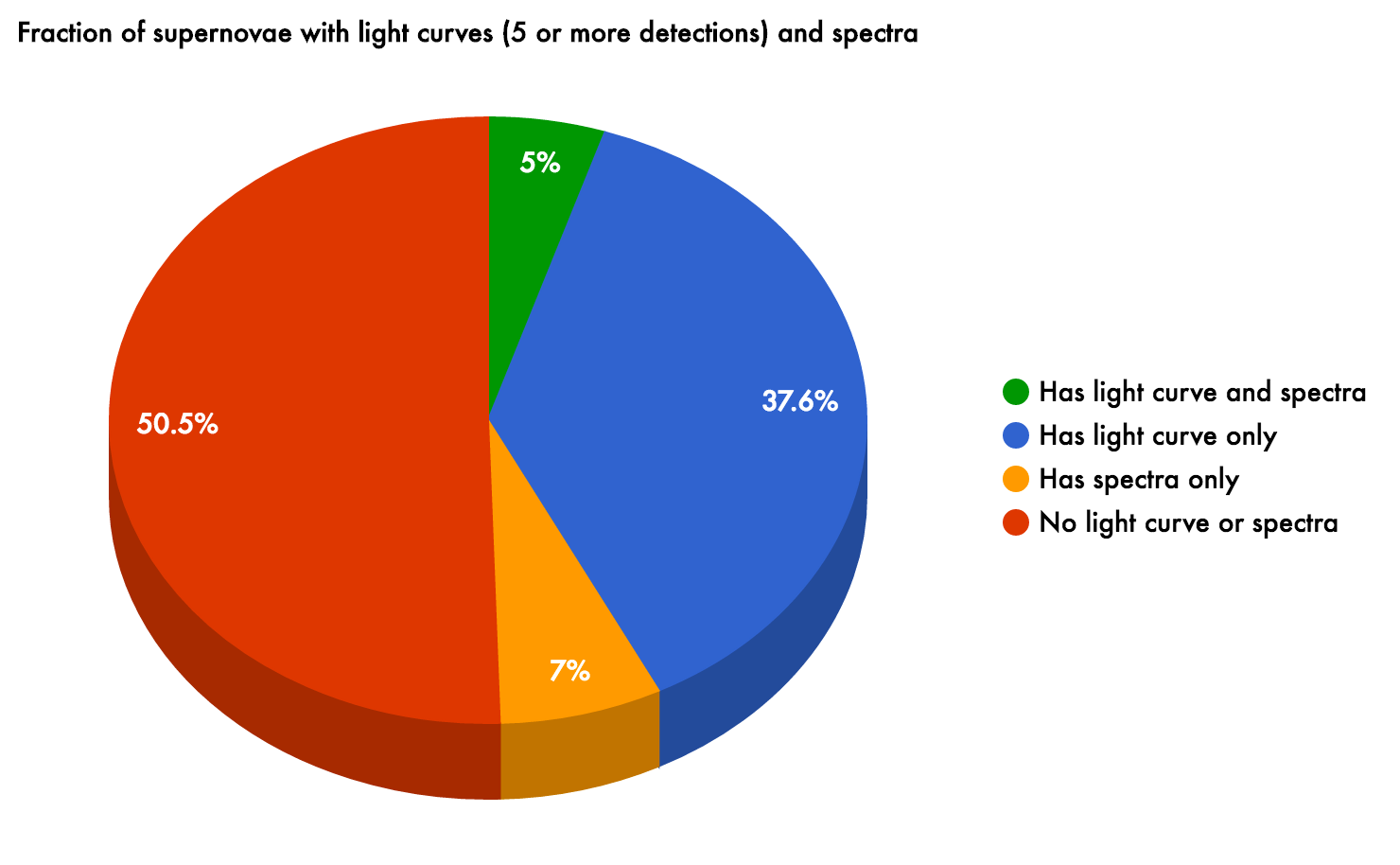}
\caption{Pie charts showing the fractions of types of events and the fractions with spectral and photometric data within the \osc. Of those events that are not classified as candidates, 31\% have at least one supernova typing, but only 12\%, one third of the typed supernovae, have at least one publicly available spectrum. An up-to-date census is available on the statistics page (\url{https://sne.space/statistics/}).}
\label{Fig:allevents}
\end{figure*}

Estimating the amount of missing data is difficult to do with the available metadata, and we can only use the data we have collected to place upper limits on the data we are missing. The numbers reported below all correspond to the \osc's state on November 7, 2016\footnote{\git commit SHA for main repository: {\tt b211cb7}}, and are displayed graphically as pie charts in Figure~\ref{Fig:allevents}.

\subsubsection{Estimate of amount of missing spectra}
At present the catalog documents 11,362 supernovae with assigned supernova types but no collected spectra, yet these supernovae almost certainly have a classification spectrum that is not available in the public domain. We have performed a cursory search of the literature to attempt to locate the publications where some of these spectra might be available (a listing of which is provided in Table~\ref{tab:missing}), a list which only identifies a small fraction of sources for the missing spectra. It is clear that the community must come together to fill this gap to adhere to basic standards of scientific reproducability and disclosure.

\subsubsection{Estimate of amount of missing light curve data}
Clues to how much available light curve data remains unaccounted for are more elusive than for spectra. There are currently 8074 supernovae with at least three but no more than ten photometric detections recorded in the catalog. We speculate that these events likely have more complete light curves that either have yet to be published or are available in published literature that we are not aware of. Additionally, there are 577 supernovae with at least one collected spectrum but no photometric observations. These too are likely to have a light curve that we have not yet collected. Finally, as mentioned above, there are over 10,000 supernovae that have been classified but have no publicly available spectrum; a large fraction of these supernovae are also likely to have light curves as decisions of which supernovae to follow up are often predicated on their early photometric evolution.

\subsection{Astroquery and VOTable support}
At present, all of the data collected by the \osc can be accessed either on an event-by-event basis or in summarized form via the main catalog page, or in bulk via downloads from \github. While we do provide a few examples scripts that process this bulk data in the main \osc repository, a standardized interface to the \osc available in a widely-used software package would further enhance accessibility and ease of data processing. Astroquery \citep{Ginsburg2016}, an affiliated package of the Astropy package which we use extensively \citep{2013A&A...558A..33A}, provides an easy-to-use, standard interface to online astronomical resources. As a community-driven open-source tool, it is straightforward to construct new ``services'' for Astroquery. As an example, the Open Exoplanet Catalogue \citep{Rein12} has added a service to Astroquery that enables the importation of exoplanet data. By adding such an interface to the \osc, we will enable users to easily incorporate the acquisition of supernova data into their code.

We can also improve access to the \osc by providing additional file formats aside from the \json format we have chosen, including the VOTable format which is \xml-based, or any other format should the need arise. The most important facet of our approach is that we have homogenized the supernova dataset to use one format; should \json fall out of fashion in the future, it will be trivial to export the dataset to another format.

\subsection{Summary Plots and Event comparisons}

Once the data have been cleaned and standardized (Section 7.1), the catalog can begin to take on comparative analysis, whereby event data can be directly compared, or sequenced, within the remaining catalog. Since a majority of published works on supernovae spectra have stemmed from visual senses (see \citealt{Parrent16,Parrent1612dn} and \citealt{Black16} for reviews), a secondary objective for the catalog is to expand our use of modern data-driven libraries, e.g., {\tt Bokeh} and {\tt d3.js}, for additional data visualization and manipulation.

Useful tools that could be incorporated into the \osc's interface are those that can be used to compare data between events, e.g. a time sequence of spectra or a collection of light curves from all events of a given type. Such tools could be used to help correct the data presented on the catalog, as is possible with the duplicate and conflict finders described in Section~\ref{sec:metadata}. While we do not intend the entirety of the scientific process to occur on the catalog itself, we hope to add simple tools that will enable simple analyses such as comparing events to one another in the near future.

\subsection{Improving and Adding Derived Quantities}

At some point the catalog can embark on
improving certain derived quantities, e.g., Milky Way Galaxy and host extinctions, 
dates of maximum brightness, and dates of explosion.
Other derived quantities that may be added in the future include line
velocities of common elements, refined subtype classifications 
(see \citealt{Benetti05,Branch06,WangX09Subtype}), estimations of photometric rise times and decline rates (e.g., $\Delta$m$_{15}$), and parameters associated with parametric fits to supernova data such as \texttt{SALT 2.4} \citep{Guy07} and \texttt{MLCS2k2} \citep{Jha07}.

As with any heterogenous collection of data, the users of the \osc must exercise caution before combining datasets without considering the systematic effects that might be present in the data that they consider. This is especially true for supernova data that is used for cosmological purposes, for which careful photometric calibration (which depends on bandpass definitions and the magnitude systems employed) and a detailed understanding of the covariance of the data are required to obtain a precise result. The \osc however provides an opportunity to encapsulate these stricter data requirements within a common format. The inclusion of additional data structures, such as covariance matrices and photometric standards, within future versions of the schema could permit cosmological datasets to be more-directly compared to one another.

\section{Summary}\label{sec:summary}

Utilizing the scattering of available observations to both vet and constrain a given explosion model (with its own blended spectrum) is not currently possible with today's large spectrum-limited survey data. There is thus a serious need for quantitative consensus in how supernovae are classified and analyzed in real time.

The accumulation of additional well-observed supernova prototypes through coordinated transient surveys, e.g., the Zwicky Transient Factory, will eventually be fueled by the Large Synoptic Survey Telescope by 2023. Taking full advantage of this surplus data stream will require extended visibility of data assets and meta-data, complete and efficient data accumulation, and full transparency from published spectra and photometry sourced from proprietary analysis tools as well.

The \osc represents a concerted effort to bring all public data of a specific kind of astrophysical transient to a single location where it can be rapidly searched, compared, and vetted. Once the data is all in a single place, analysis is significantly simplified for the user, which reduces the cumulative time expenditure of the astrophysical community which has repeated the collection process many times. Other astrophysical datasets could also benefit from the system we have presented here, including other transient phenomena such as novae, cataclysmic variables, gamma ray bursts, and tidal disruption events\footnote{\url{https://tde.space}}.

\acknowledgments

We thank Brad~Cenko, Maria~Drout, Or~Graur, Atish~Kamble, Bob~Kirschner, Dan~Milisavljevic, Gautham~Narayan, Matt~Nicholl, Ashley~Pagnotta, Ed~Shaya, Isaac~Shivvers, Jeff~Silverman, Alicia~Soderberg, and {\L}ukasz Wyrzykowski for constructive suggestions and comments. The authors are especially grateful to Ryan~Foley for providing user interface feedback; to Dmitry~Tsvetkov and Pavlyuk~Nikolay who donated hundreds of historical light curves; to Pete Challis, Koji Kawabata, Hagai Perets, and Federica Bianco for providing a significant number of spectra; to Matt~Nicholl and Llu\'{\i}s Galbany for donating their personal supernova light curve collections; to Carles~Badenes, Dave Green, and Pierre~Maggi for their SNR contributions; to Dan~Milisavljevic for testing the duplicate finder (from data to appear in Milisavljevic et al., in prep.); to Federica~Bianco, Matt~Nicholl, Maria~Pruzhinskaya, and Victor~Lebedev (who also pointed us to many missing supernovae), and Ofer~Yaron for finding several errors in the catalog; and to David~Bishop for his close cooperation and tireless effort to maintain the Latest Supernovae webpage\footnote{\url{http://www.rochesterastronomy.org/supernova.html}}. The authors would also like to thank the anonymous referee for their constructive comments.

This work was supported in part by Einstein grant PF3-140108 (J. G.). The \osc made use of NASA's Astrophysics Data System; the VizieR catalogue access tool, CDS, Strasbourg, France; Astropy, a community-developed core \python package for Astronomy \citep{2013A&A...558A..33A}; Astroquery \citep{Ginsburg2016}, Beautiful Soup\footnote{\url{https://www.crummy.com/software/BeautifulSoup/bs4/doc/}}; DataTables\footnote{\url{https://datatables.net/}}; the Wordpress blogging platform\footnote{\url{http://wordpress.com/}}; the Bokeh interactive visualization library\footnote{\url{http://bokeh.pydata.org/en/latest/}}; and \mbox{DownloadThemAll}, a Firefox add-on\footnote{\url{http://www.downthemall.net/}}.

\bibliographystyle{aasjournal}
\bibliography{jparrent_bib}

\begin{thebibliography}{}
\expandafter\ifx\csname natexlab\endcsname\relax\def\natexlab#1{#1}\fi

\bibitem[{{Aldering} {et~al.}(2002){Aldering}, {Adam}, {Antilogus}, {Astier},
  {Bacon}, {Bongard}, {Bonnaud}, {Copin}, {Hardin}, {Henault}, {Howell},
  {Lemonnier}, {Levy}, {Loken}, {Nugent}, {Pain}, {Pecontal}, {Pecontal},
  {Perlmutter}, {Quimby}, {Schahmaneche}, {Smadja}, \&
  {Wood-Vasey}}]{Aldering02}
{Aldering}, G., {Adam}, G., {Antilogus}, P., {et~al.} 2002, in Society of
  Photo-Optical Instrumentation Engineers (SPIE) Conference Series, Vol. 4836,
  Society of Photo-Optical Instrumentation Engineers (SPIE) Conference Series,
  ed. J.~A. {Tyson} \& S.~{Wolff}, 61--72

\bibitem[{{Andrews} {et~al.}(2011){Andrews}, {Sugerman}, {Clayton},
  {Gallagher}, {Barlow}, {Clem}, {Ercolano}, {Fabbri}, {Meixner}, {Otsuka},
  {Welch}, \& {Wesson}}]{Andrews11}
{Andrews}, J.~E., {Sugerman}, B.~E.~K., {Clayton}, G.~C., {et~al.} 2011, \apj,
  731, 47

\bibitem[{{Anupama}(1997)}]{Anupama97}
{Anupama}, G.~C. 1997, \aj, 114, 2054

\bibitem[{{Astropy Collaboration} {et~al.}(2013){Astropy Collaboration},
  {Robitaille}, {Tollerud}, {Greenfield}, {Droettboom}, {Bray}, {Aldcroft},
  {Davis}, {Ginsburg}, {Price-Whelan}, {Kerzendorf}, {Conley}, {Crighton},
  {Barbary}, {Muna}, {Ferguson}, {Grollier}, {Parikh}, {Nair}, {Unther},
  {Deil}, {Woillez}, {Conseil}, {Kramer}, {Turner}, {Singer}, {Fox}, {Weaver},
  {Zabalza}, {Edwards}, {Azalee Bostroem}, {Burke}, {Casey}, {Crawford},
  {Dencheva}, {Ely}, {Jenness}, {Labrie}, {Lim}, {Pierfederici}, {Pontzen},
  {Ptak}, {Refsdal}, {Servillat}, \& {Streicher}}]{2013A&A...558A..33A}
{Astropy Collaboration}, {Robitaille}, T.~P., {Tollerud}, E.~J., {et~al.} 2013,
  \aap, 558, A33

\bibitem[{{Balland} {et~al.}(2009){Balland}, {Baumont}, {Basa}, {Mouchet},
  {Howell}, {Astier}, {Carlberg}, {Conley}, {Fouchez}, {Guy}, {Hardin}, {Hook},
  {Pain}, {Perrett}, {Pritchet}, {Regnault}, {Rich}, {Sullivan}, {Antilogus},
  {Arsenijevic}, {Le Du}, {Fabbro}, {Lidman}, {Mour{\~a}o},
  {Palanque-Delabrouille}, {P{\'e}contal}, \& {Ruhlmann-Kleider}}]{Balland09}
{Balland}, C., {Baumont}, S., {Basa}, S., {et~al.} 2009, A\&A, 507, 85

\bibitem[{{Baltay} {et~al.}(2012){Baltay}, {Rabinowitz}, {Hadjiyska},
  {Schwamb}, {Ellman}, {Zinn}, {Tourtellotte}, {McKinnon}, {Horowitz},
  {Effron}, \& {Nugent}}]{Baltay12}
{Baltay}, C., {Rabinowitz}, D., {Hadjiyska}, E., {et~al.} 2012, The Messenger,
  150, 34

\bibitem[{{Barbon} {et~al.}(1999){Barbon}, {Buond{\'{\i}}}, {Cappellaro}, \&
  {Turatto}}]{Barbon99}
{Barbon}, R., {Buond{\'{\i}}}, V., {Cappellaro}, E., \& {Turatto}, M. 1999,
  A\&AS, 139, 531

\bibitem[{{Ben-Ami} {et~al.}(2014){Ben-Ami}, {Gal-Yam}, {Mazzali}, {Gnat},
  {Modjaz}, {Rabinak}, {Sullivan}, {Bildsten}, {Poznanski}, {Yaron}, {Arcavi},
  {Bloom}, {Horesh}, {Kasliwal}, {Kulkarni}, {Nugent}, {Ofek}, {Perley},
  {Quimby}, \& {Xu}}]{BenAmi14}
{Ben-Ami}, S., {Gal-Yam}, A., {Mazzali}, P.~A., {et~al.} 2014, \apj, 785, 37

\bibitem[{{Ben-Ami} {et~al.}(2015){Ben-Ami}, {Hachinger}, {Gal-Yam}, {Mazzali},
  {Filippenko}, {Horesh}, {Matheson}, {Modjaz}, {Sauer}, {Silverman}, {Smith},
  \& {Yaron}}]{BenAmi15}
{Ben-Ami}, S., {Hachinger}, S., {Gal-Yam}, A., {et~al.} 2015, \apj, 803, 40

\bibitem[{{Benetti} {et~al.}(2002){Benetti}, {Branch}, {Turatto}, {Cappellaro},
  {Baron}, {Zampieri}, {Della Valle}, \& {Pastorello}}]{Benetti02}
{Benetti}, S., {Branch}, D., {Turatto}, M., {et~al.} 2002, \mnras, 336, 91

\bibitem[{{Benetti} {et~al.}(2005){Benetti}, {Cappellaro}, {Mazzali},
  {Turatto}, {Altavilla}, {Bufano}, {Elias-Rosa}, {Kotak}, {Pignata}, {Salvo},
  \& {Stanishev}}]{Benetti05}
{Benetti}, S., {Cappellaro}, E., {Mazzali}, P.~A., {et~al.} 2005, ApJ, 623,
  1011

\bibitem[{{Black} {et~al.}(2016){Black}, {Fesen}, \& {Parrent}}]{Black16}
{Black}, C.~S., {Fesen}, R.~A., \& {Parrent}, J.~T. 2016, ArXiv e-prints,
  arXiv:1604.01044

\bibitem[{{Blondin} {et~al.}(2012){Blondin}, {Matheson}, {Kirshner}, {Mandel},
  {Berlind}, {Calkins}, {Challis}, {Garnavich}, {Jha}, {Modjaz}, {Riess}, \&
  {Schmidt}}]{Blondin12}
{Blondin}, S., {Matheson}, T., {Kirshner}, R.~P., {et~al.} 2012, AJ, 143, 126

\bibitem[{{Bloom} {et~al.}(2012){Bloom}, {Kasen}, {Shen}, {Nugent}, {Butler},
  {Graham}, {Howell}, {Kolb}, {Holmes}, {Haswell}, {Burwitz}, {Rodriguez}, \&
  {Sullivan}}]{Bloom12}
{Bloom}, J.~S., {Kasen}, D., {Shen}, K.~J., {et~al.} 2012, ApJL, 744, L17

\bibitem[{{Bose} {et~al.}(2015){Bose}, {Valenti}, {Misra}, {Pumo}, {Zampieri},
  {Sand}, {Kumar}, {Pastorello}, {Sutaria}, {Maccarone}, {Kumar}, {Graham},
  {Howell}, {Ochner}, {Chandola}, \& {Pandey}}]{Bose15}
{Bose}, S., {Valenti}, S., {Misra}, K., {et~al.} 2015, \mnras, 450, 2373

\bibitem[{{Botticella} {et~al.}(2010){Botticella}, {Trundle}, {Pastorello},
  {Rodney}, {Rest}, {Gezari}, {Smartt}, {Narayan}, {Huber}, {Tonry}, {Young},
  {Smith}, {Bresolin}, {Valenti}, {Kotak}, {Mattila}, {Kankare}, {Wood-Vasey},
  {Riess}, {Neill}, {Forster}, {Martin}, {Stubbs}, {Burgett}, {Chambers},
  {Dombeck}, {Flewelling}, {Grav}, {Heasley}, {Hodapp}, {Kaiser}, {Kudritzki},
  {Luppino}, {Lupton}, {Magnier}, {Monet}, {Morgan}, {Onaka}, {Price},
  {Rhoads}, {Siegmund}, {Sweeney}, {Wainscoat}, {Waters}, {Waterson}, \&
  {Wynn-Williams}}]{Botticella10}
{Botticella}, M.~T., {Trundle}, C., {Pastorello}, A., {et~al.} 2010, \apjl,
  717, L52

\bibitem[{{Branch} {et~al.}(2005){Branch}, {Baron}, {Hall}, {Melakayil}, \&
  {Parrent}}]{Branch05}
{Branch}, D., {Baron}, E., {Hall}, N., {Melakayil}, M., \& {Parrent}, J. 2005,
  PASP, 117, 545

\bibitem[{{Branch} {et~al.}(1982){Branch}, {Buta}, {Falk}, {McCall}, {Uomoto},
  {Wheeler}, {Wills}, \& {Sutherland}}]{Branch82}
{Branch}, D., {Buta}, R., {Falk}, S.~W., {et~al.} 1982, ApJL, 252, L61

\bibitem[{{Branch} {et~al.}(2009){Branch}, {Dang}, \& {Baron}}]{Branch09}
{Branch}, D., {Dang}, L.~C., \& {Baron}, E. 2009, PASP, 121, 238

\bibitem[{{Branch} {et~al.}(2006){Branch}, {Dang}, {Hall}, {Ketchum},
  {Melakayil}, {Parrent}, {Troxel}, {Casebeer}, {Jeffery}, \&
  {Baron}}]{Branch06}
{Branch}, D., {Dang}, L.~C., {Hall}, N., {et~al.} 2006, PASP, 118, 560

\bibitem[{{Branch} {et~al.}(2007){Branch}, {Troxel}, {Jeffery}, {Hatano},
  {Musco}, {Parrent}, {Baron}, {Dang}, {Casebeer}, {Hall}, \&
  {Ketchum}}]{Branch07b}
{Branch}, D., {Troxel}, M.~A., {Jeffery}, D.~J., {et~al.} 2007, PASP, 119, 709

\bibitem[{{Branch} {et~al.}(2008){Branch}, {Jeffery}, {Parrent}, {Baron},
  {Troxel}, {Stanishev}, {Keithley}, {Harrison}, \& {Bruner}}]{Branch08}
{Branch}, D., {Jeffery}, D.~J., {Parrent}, J., {et~al.} 2008, PASP, 120, 135

\bibitem[{{Brown} {et~al.}(2005){Brown}, {Holland}, {James}, {Milne}, {Roming},
  {Mason}, {Page}, {Beardmore}, {Burrows}, {Morgan}, {Gronwall}, {Blustin},
  {Boyd}, {Still}, {Breeveld}, {de Pasquale}, {Hunsberger}, {Ivanushkina},
  {Landsman}, {McGowan}, {Poole}, {Rosen}, {Schady}, \& {Gehrels}}]{Brown05}
{Brown}, P.~J., {Holland}, S.~T., {James}, C., {et~al.} 2005, ApJ, 635, 1192

\bibitem[{{Campana} {et~al.}(2006){Campana}, {Mangano}, {Blustin}, {Brown},
  {Burrows}, {Chincarini}, {Cummings}, {Cusumano}, {Della Valle}, {Malesani},
  {M{\'e}sz{\'a}ros}, {Nousek}, {Page}, {Sakamoto}, {Waxman}, {Zhang}, {Dai},
  {Gehrels}, {Immler}, {Marshall}, {Mason}, {Moretti}, {O'Brien}, {Osborne},
  {Page}, {Romano}, {Roming}, {Tagliaferri}, {Cominsky}, {Giommi}, {Godet},
  {Kennea}, {Krimm}, {Angelini}, {Barthelmy}, {Boyd}, {Palmer}, {Wells}, \&
  {White}}]{Campana06}
{Campana}, S., {Mangano}, V., {Blustin}, A.~J., {et~al.} 2006, \nat, 442, 1008

\bibitem[{{Cano} {et~al.}(2014){Cano}, {de Ugarte Postigo}, {Pozanenko},
  {Butler}, {Th{\"o}ne}, {Guidorzi}, {Kr{\"u}hler}, {Gorosabel}, {Jakobsson},
  {Leloudas}, {Malesani}, {Hjorth}, {Melandri}, {Mundell}, {Wiersema},
  {D'Avanzo}, {Schulze}, {Gomboc}, {Johansson}, {Zheng}, {Kann}, {Knust},
  {Varela}, {Akerlof}, {Bloom}, {Burkhonov}, {Cooke}, {de Diego}, {Dhungana},
  {Farina}, {Ferrante}, {Flewelling}, {Fox}, {Fynbo}, {Gehrels}, {Georgiev},
  {Gonz{\'a}lez}, {Greiner}, {G{\"u}ver}, {Hartoog}, {Hatch}, {Jelinek},
  {Kehoe}, {Klose}, {Klunko}, {Kopa{\v c}}, {Kutyrev}, {Krugly}, {Lee},
  {Levan}, {Linkov}, {Matkin}, {Minikulov}, {Molotov}, {Prochaska}, {Richer},
  {Rom{\'a}n-Z{\'u}{\~n}iga}, {Rumyantsev}, {S{\'a}nchez-Ram{\'{\i}}rez},
  {Steele}, {Tanvir}, {Volnova}, {Watson}, {Xu}, \& {Yuan}}]{Cano1413fu}
{Cano}, Z., {de Ugarte Postigo}, A., {Pozanenko}, A., {et~al.} 2014, \aap, 568,
  A19

\bibitem[{{Cao} {et~al.}(2015){Cao}, {Kulkarni}, {Howell}, {Gal-Yam},
  {Kasliwal}, {Valenti}, {Johansson}, {Amanullah}, {Goobar}, {Sollerman},
  {Taddia}, {Horesh}, {Sagiv}, {Cenko}, {Nugent}, {Arcavi}, {Surace},
  {Wo{\'z}niak}, {Moody}, {Rebbapragada}, {Bue}, \& {Gehrels}}]{Cao15}
{Cao}, Y., {Kulkarni}, S.~R., {Howell}, D.~A., {et~al.} 2015, \nat, 521, 328

\bibitem[{{Cao} {et~al.}(2016){Cao}, {Johansson}, {Nugent}, {Goobar}, {Nordin},
  {Kulkarni}, {Cenko}, {Fox}, {Kasliwal}, {Fremling}, {Amanullah}, {Hsiao},
  {Perley}, {Bue}, {Masci}, {Lee}, \& {Chotard}}]{Cao16}
{Cao}, Y., {Johansson}, J., {Nugent}, P.~E., {et~al.} 2016, ArXiv e-prints,
  arXiv:1601.00686

\bibitem[{{Cartier} {et~al.}(2014){Cartier}, {Hamuy}, {Pignata}, {F{\"o}rster},
  {Zelaya}, {Folatelli}, {Phillips}, {Morrell}, {Krisciunas}, {Suntzeff},
  {Clocchiatti}, {Coppi}, {Contreras}, {Roth}, {Koviak}, {Maza},
  {Gonz{\'a}lez}, {Gonz{\'a}lez}, \& {Huerta}}]{Cartier14}
{Cartier}, R., {Hamuy}, M., {Pignata}, G., {et~al.} 2014, ApJ, 789, 89

\bibitem[{{Casebeer} {et~al.}(1998){Casebeer}, {Blaylock}, {Deaton}, {Branch},
  {Baron}, {Richardson}, \& {Ancheta}}]{Casebeer98}
{Casebeer}, D., {Blaylock}, M., {Deaton}, J., {et~al.} 1998, in Bulletin of the
  American Astronomical Society, Vol.~30, American Astronomical Society Meeting
  Abstracts, 1324

\bibitem[{{Chakradhari} {et~al.}(2014){Chakradhari}, {Sahu}, {Srivastav}, \&
  {Anupama}}]{Chakradhari14}
{Chakradhari}, N.~K., {Sahu}, D.~K., {Srivastav}, S., \& {Anupama}, G.~C. 2014,
  MNRAS, 443, 1663

\bibitem[{{Chen} {et~al.}(2014){Chen}, {Wang}, {Ganeshalingam}, {Silverman},
  {Filippenko}, {Li}, {Chornock}, {Li}, \& {Steele}}]{Chen14}
{Chen}, J., {Wang}, X., {Ganeshalingam}, M., {et~al.} 2014, \apj, 790, 120

\bibitem[{{Chornock} {et~al.}(2013){Chornock}, {Berger}, {Rest},
  {Milisavljevic}, {Lunnan}, {Foley}, {Soderberg}, {Smartt}, {Burgasser},
  {Challis}, {Chomiuk}, {Czekala}, {Drout}, {Fong}, {Huber}, {Kirshner},
  {Leibler}, {McLeod}, {Marion}, {Narayan}, {Riess}, {Roth}, {Sanders},
  {Scolnic}, {Smith}, {Stubbs}, {Tonry}, {Valenti}, {Burgett}, {Chambers},
  {Hodapp}, {Kaiser}, {Kudritzki}, {Magnier}, \& {Price}}]{Chornock13}
{Chornock}, R., {Berger}, E., {Rest}, A., {et~al.} 2013, \apj, 767, 162

\bibitem[{{Clark} \& {Caswell}(1976)}]{Clark76}
{Clark}, D.~H., \& {Caswell}, J.~L. 1976, \mnras, 174, 267

\bibitem[{{Clocchiatti} {et~al.}(1997){Clocchiatti}, {Wheeler}, {Phillips},
  {Suntzeff}, {Cristiani}, {Phillips}, {Harkness}, {Dopita}, {Beuermann},
  {Rosa}, {Grosb{\o}l}, {Lindblad}, \& {Filippenko}}]{Clocchiatti97}
{Clocchiatti}, A., {Wheeler}, J.~C., {Phillips}, M.~M., {et~al.} 1997, \apj,
  483, 675

\bibitem[{{Clocchiatti} {et~al.}(2000){Clocchiatti}, {Phillips}, {Suntzeff},
  {Della Valle}, {Cappellaro}, {Turatto}, {Hamuy}, {Avil{\'e}s}, {Navarrete},
  {Smith}, {Rubenstein}, {Covarrubias}, {Stetson}, {Maza}, {Riess}, \&
  {Zanin}}]{Clocchiatti00}
{Clocchiatti}, A., {Phillips}, M.~M., {Suntzeff}, N.~B., {et~al.} 2000, \apj,
  529, 661

\bibitem[{{Di Carlo} {et~al.}(2002){Di Carlo}, {Massi}, {Valentini}, {Di
  Paola}, {D'Alessio}, {Brocato}, {Guidubaldi}, {Dolci}, {Pedichini},
  {Speziali}, {Li Causi}, {Caratti o Garatti}, {Cappellaro}, {Turatto},
  {Arkharov}, {Gnedin}, {Larionov}, {Benetti}, {Pastorello}, {Aretxaga},
  {Chavushyan}, {Vega}, {Danziger}, \& {Tornamb{\'e}}}]{DiCarlo02}
{Di Carlo}, E., {Massi}, F., {Valentini}, G., {et~al.} 2002, \apj, 573, 144

\bibitem[{{Diamond} {et~al.}(2015){Diamond}, {Hoeflich}, \&
  {Gerardy}}]{Diamond15}
{Diamond}, T.~R., {Hoeflich}, P., \& {Gerardy}, C.~L. 2015, \apj, 806, 107

\bibitem[{{Drake} {et~al.}(2010){Drake}, {Djorgovski}, {Prieto}, {Mahabal},
  {Balam}, {Williams}, {Graham}, {Catelan}, {Beshore}, \& {Larson}}]{Drake10}
{Drake}, A.~J., {Djorgovski}, S.~G., {Prieto}, J.~L., {et~al.} 2010, \apjl,
  718, L127

\bibitem[{{Elias-Rosa} {et~al.}(2011){Elias-Rosa}, {Van Dyk}, {Li},
  {Silverman}, {Foley}, {Ganeshalingam}, {Mauerhan}, {Kankare}, {Jha},
  {Filippenko}, {Beckman}, {Berger}, {Cuillandre}, \& {Smith}}]{EliasRosa11}
{Elias-Rosa}, N., {Van Dyk}, S.~D., {Li}, W., {et~al.} 2011, \apj, 742, 6

\bibitem[{{Elmhamdi} {et~al.}(2003){Elmhamdi}, {Danziger}, {Chugai},
  {Pastorello}, {Turatto}, {Cappellaro}, {Altavilla}, {Benetti}, {Patat}, \&
  {Salvo}}]{Elmhamdi03}
{Elmhamdi}, A., {Danziger}, I.~J., {Chugai}, N., {et~al.} 2003, \mnras, 338,
  939

\bibitem[{{Falk} \& {Arnett}(1977)}]{Falk77}
{Falk}, S.~W., \& {Arnett}, W.~D. 1977, \apjs, 33, 515

\bibitem[{{Fassia} {et~al.}(1998){Fassia}, {Meikle}, {Geballe}, {Walton},
  {Pollacco}, {Rutten}, \& {Tinney}}]{Fassia98}
{Fassia}, A., {Meikle}, W.~P.~S., {Geballe}, T.~R., {et~al.} 1998, \mnras, 299,
  150

\bibitem[{{Flin} {et~al.}(1979){Flin}, {Karpowicz}, {Murawski}, \&
  {Rudnicki}}]{FLIN79}
{Flin}, P., {Karpowicz}, M., {Murawski}, W., \& {Rudnicki}, K. 1979, Acta
  Cosmologica, 8, 5

\bibitem[{{Folatelli} {et~al.}(2013){Folatelli}, {Morrell}, {Phillips},
  {Hsiao}, {Campillay}, {Contreras}, {Castell{\'o}n}, {Hamuy}, {Krzeminski},
  {Roth}, {Stritzinger}, {Burns}, {Freedman}, {Madore}, {Murphy}, {Persson},
  {Prieto}, {Suntzeff}, {Krisciunas}, {Anderson}, {F{\"o}rster}, {Maza},
  {Pignata}, {Rojas}, {Boldt}, {Salgado}, {Wyatt}, {Olivares E.}, {Gal-Yam}, \&
  {Sako}}]{Folatelli13}
{Folatelli}, G., {Morrell}, N., {Phillips}, M.~M., {et~al.} 2013, ApJ, 773, 53

\bibitem[{{Foley} {et~al.}(2010){Foley}, {Rest}, {Stritzinger}, {Pignata},
  {Anderson}, {Hamuy}, {Morrell}, {Phillips}, \& {Salgado}}]{Foley1008ge}
{Foley}, R.~J., {Rest}, A., {Stritzinger}, M., {et~al.} 2010, \aj, 140, 1321

\bibitem[{{Foley} {et~al.}(2012){Foley}, {Challis}, {Filippenko},
  {Ganeshalingam}, {Landsman}, {Li}, {Marion}, {Silverman}, {Beaton},
  {Bennert}, {Cenko}, {Childress}, {Guhathakurta}, {Jiang}, {Kalirai},
  {Kirshner}, {Stockton}, {Tollerud}, {Vink{\'o}}, {Wheeler}, \&
  {Woo}}]{Foley12c}
{Foley}, R.~J., {Challis}, P.~J., {Filippenko}, A.~V., {et~al.} 2012, ApJ, 744,
  38

\bibitem[{{Fransson} {et~al.}(2002){Fransson}, {Chevalier}, {Filippenko},
  {Leibundgut}, {Barth}, {Fesen}, {Kirshner}, {Leonard}, {Li}, {Lundqvist},
  {Sollerman}, \& {Van Dyk}}]{Fransson02}
{Fransson}, C., {Chevalier}, R.~A., {Filippenko}, A.~V., {et~al.} 2002, \apj,
  572, 350

\bibitem[{{Fransson} {et~al.}(2014){Fransson}, {Ergon}, {Challis}, {Chevalier},
  {France}, {Kirshner}, {Marion}, {Milisavljevic}, {Smith}, {Bufano},
  {Friedman}, {Kangas}, {Larsson}, {Mattila}, {Benetti}, {Chornock}, {Czekala},
  {Soderberg}, \& {Sollerman}}]{Fransson14}
{Fransson}, C., {Ergon}, M., {Challis}, P.~J., {et~al.} 2014, \apj, 797, 118

\bibitem[{{Fraser} {et~al.}(2010){Fraser}, {Tak{\'a}ts}, {Pastorello},
  {Smartt}, {Mattila}, {Botticella}, {Valenti}, {Ergon}, {Sollerman}, {Arcavi},
  {Benetti}, {Bufano}, {Crockett}, {Danziger}, {Gal-Yam}, {Maund},
  {Taubenberger}, \& {Turatto}}]{Fraser10}
{Fraser}, M., {Tak{\'a}ts}, K., {Pastorello}, A., {et~al.} 2010, \apjl, 714,
  L280

\bibitem[{{Gallagher} {et~al.}(2012){Gallagher}, {Sugerman}, {Clayton},
  {Andrews}, {Clem}, {Barlow}, {Ercolano}, {Fabbri}, {Otsuka}, {Wesson}, \&
  {Meixner}}]{Gallagher12}
{Gallagher}, J.~S., {Sugerman}, B.~E.~K., {Clayton}, G.~C., {et~al.} 2012,
  \apj, 753, 109

\bibitem[{{Garnavich} {et~al.}(2016){Garnavich}, {Tucker}, {Rest}, {Shaya},
  {Olling}, {Kasen}, \& {Villar}}]{Garnavich16}
{Garnavich}, P.~M., {Tucker}, B.~E., {Rest}, A., {et~al.} 2016, \apj, 820, 23

\bibitem[{{Gerardy} {et~al.}(2002){Gerardy}, {Fesen}, {Nomoto}, {Maeda},
  {Hoflich}, \& {Wheeler}}]{Gerardy02}
{Gerardy}, C.~L., {Fesen}, R.~A., {Nomoto}, K., {et~al.} 2002, \pasj, 54, 905

\bibitem[{{Gezari} {et~al.}(2010){Gezari}, {Rest}, {Huber}, {Narayan},
  {Forster}, {Neill}, {Martin}, {Valenti}, {Smartt}, {Chornock}, {Berger},
  {Soderberg}, {Mattila}, {Kankare}, {Burgett}, {Chambers}, {Dombeck}, {Grav},
  {Heasley}, {Hodapp}, {Jedicke}, {Kaiser}, {Kudritzki}, {Luppino}, {Lupton},
  {Magnier}, {Monet}, {Morgan}, {Onaka}, {Price}, {Rhoads}, {Siegmund},
  {Stubbs}, {Tonry}, {Wainscoat}, {Waterson}, \& {Wynn-Williams}}]{Gezari10}
{Gezari}, S., {Rest}, A., {Huber}, M.~E., {et~al.} 2010, \apjl, 720, L77

\bibitem[{{Gezari} {et~al.}(2015){Gezari}, {Jones}, {Sanders}, {Soderberg},
  {Hung}, {Heinis}, {Smartt}, {Rest}, {Scolnic}, {Chornock}, {Berger}, {Foley},
  {Huber}, {Price}, {Stubbs}, {Riess}, {Kirshner}, {Smith}, {Wood-Vasey},
  {Schiminovich}, {Martin}, {Burgett}, {Chambers}, {Flewelling}, {Kaiser},
  {Tonry}, \& {Wainscoat}}]{Gezari15}
{Gezari}, S., {Jones}, D.~O., {Sanders}, N.~E., {et~al.} 2015, \apj, 804, 28

\bibitem[{Ginsburg {et~al.}(2016)Ginsburg, Parikh, Woillez, Groener, Sipocz,
  Liedtke, Robitaille, Deil, Svoboda, Tollerud, Persson, adamginsburg,
  Séguin-Charbonneau, Armstrong, Mirocha, Droettboom, james allen, Moolekamp,
  Egeland, Singer, Barbary, Grollier, Shiga, Günther, Booker, Parejko, Rol,
  Edward, Miller, \& Fragodt}]{Ginsburg2016}
Ginsburg, A., Parikh, M., Woillez, J., {et~al.} 2016, astroquery: v0.3.2
  release, , , doi:10.5281/zenodo.55253

\bibitem[{{Goobar} {et~al.}(2014){Goobar}, {Johansson}, {Amanullah}, {Cao},
  {Perley}, {Kasliwal}, {Ferretti}, {Nugent}, {Harris}, {Gal-Yam}, {Ofek},
  {Tendulkar}, {Dennefeld}, {Valenti}, {Arcavi}, {Banerjee}, {Venkataraman},
  {Joshi}, {Ashok}, {Cenko}, {Diaz}, {Fremling}, {Horesh}, {Howell},
  {Kulkarni}, {Papadogiannakis}, {Petrushevska}, {Sand}, {Sollerman},
  {Stanishev}, {Bloom}, {Surace}, {Dupuy}, \& {Liu}}]{Goobar14}
{Goobar}, A., {Johansson}, J., {Amanullah}, R., {et~al.} 2014, \apjl, 784, L12

\bibitem[{{Graham} {et~al.}(2015){Graham}, {Foley}, {Zheng}, {Kelly},
  {Shivvers}, {Silverman}, {Filippenko}, {Clubb}, \&
  {Ganeshalingam}}]{Graham15}
{Graham}, M.~L., {Foley}, R.~J., {Zheng}, W., {et~al.} 2015, MNRAS, 446, 2073

\bibitem[{{Graur} {et~al.}(2016{\natexlab{a}}){Graur}, {Bianco}, {Huang},
  {Modjaz}, {Shivvers}, {Filippenko}, \& {Li}}]{Graur:2016b}
{Graur}, O., {Bianco}, F.~B., {Huang}, S., {et~al.} 2016{\natexlab{a}}, ArXiv
  e-prints, arXiv:1609.02921

\bibitem[{{Graur} {et~al.}(2016{\natexlab{b}}){Graur}, {Bianco}, {Modjaz},
  {Shivvers}, {Filippenko}, {Li}, \& {Smith}}]{Graur:2016a}
{Graur}, O., {Bianco}, F.~B., {Modjaz}, M., {et~al.} 2016{\natexlab{b}}, ArXiv
  e-prints, arXiv:1609.02923

\bibitem[{{Green}(1984)}]{Green84}
{Green}, D.~A. 1984, \mnras, 209, 449

\bibitem[{{Green}(1988)}]{Green88}
---. 1988, \apss, 148, 3

\bibitem[{{Green}(2014)}]{Green:2014a}
---. 2014, Bulletin of the Astronomical Society of India, 42, 47

\bibitem[{{Greiner} {et~al.}(2015){Greiner}, {Mazzali}, {Kann}, {Kr{\"u}hler},
  {Pian}, {Prentice}, {Olivares E.}, {Rossi}, {Klose}, {Taubenberger}, {Knust},
  {Afonso}, {Ashall}, {Bolmer}, {Delvaux}, {Diehl}, {Elliott}, {Filgas},
  {Fynbo}, {Graham}, {Guelbenzu}, {Kobayashi}, {Leloudas}, {Savaglio},
  {Schady}, {Schmidl}, {Schweyer}, {Sudilovsky}, {Tanga}, {Updike}, {van
  Eerten}, \& {Varela}}]{Greiner15}
{Greiner}, J., {Mazzali}, P.~A., {Kann}, D.~A., {et~al.} 2015, \nat, 523, 189

\bibitem[{{Guti{\'e}rrez} {et~al.}(2016){Guti{\'e}rrez},
  {Gonz{\'a}lez-Gait{\'a}n}, {Folatelli}, {Pignata}, {Anderson}, {Hamuy},
  {Morrell}, {Stritzinger}, {Taubenberger}, {Bufano}, {Olivares}, {Haislip}, \&
  {Reichart}}]{Gutierrez16}
{Guti{\'e}rrez}, C.~P., {Gonz{\'a}lez-Gait{\'a}n}, S., {Folatelli}, G.,
  {et~al.} 2016, ArXiv e-prints, arXiv:1601.07863

\bibitem[{{Guy} {et~al.}(2007){Guy}, {Astier}, {Baumont}, {Hardin}, {Pain},
  {Regnault}, {Basa}, {Carlberg}, {Conley}, {Fabbro}, {Fouchez}, {Hook},
  {Howell}, {Perrett}, {Pritchet}, {Rich}, {Sullivan}, {Antilogus}, {Aubourg},
  {Bazin}, {Bronder}, {Filiol}, {Palanque-Delabrouille}, {Ripoche}, \&
  {Ruhlmann-Kleider}}]{Guy07}
{Guy}, J., {Astier}, P., {Baumont}, S., {et~al.} 2007, A\&A, 466, 11

\bibitem[{{Hernandez} {et~al.}(2000){Hernandez}, {Meikle}, {Aparicio}, {Benn},
  {Burleigh}, {Chrysostomou}, {Fernandes}, {Geballe}, {Hammersley},
  {Iglesias-Paramo}, {James}, {James}, {Kemp}, {Lister}, {Martinez-Delgado},
  {Oscoz}, {Pollacco}, {Rozas}, {Smartt}, {Sorensen}, {Swaters}, {Telting},
  {Vacca}, {Walton}, \& {Zapatero-Osorio}}]{Hernandez00}
{Hernandez}, M., {Meikle}, W.~P.~S., {Aparicio}, A., {et~al.} 2000, MNRAS, 319,
  223

\bibitem[{{H{\"o}flich} {et~al.}(2002){H{\"o}flich}, {Gerardy}, {Fesen}, \&
  {Sakai}}]{Hoflich02}
{H{\"o}flich}, P., {Gerardy}, C.~L., {Fesen}, R.~A., \& {Sakai}, S. 2002, ApJ,
  568, 791

\bibitem[{{Huang} {et~al.}(2015){Huang}, {Wang}, {Zhang}, {Brown}, {Zampieri},
  {Pumo}, {Zhang}, {Chen}, {Mo}, \& {Zhao}}]{Huang15}
{Huang}, F., {Wang}, X., {Zhang}, J., {et~al.} 2015, \apj, 807, 59

\bibitem[{{Hunter} {et~al.}(2009){Hunter}, {Valenti}, {Kotak}, {Meikle},
  {Taubenberger}, {Pastorello}, {Benetti}, {Stanishev}, {Smartt}, {Trundle},
  {Arkharov}, {Bufano}, {Cappellaro}, {Di Carlo}, {Dolci}, {Elias-Rosa},
  {Frandsen}, {Fynbo}, {Hopp}, {Larionov}, {Laursen}, {Mazzali}, {Navasardyan},
  {Ries}, {Riffeser}, {Rizzi}, {Tsvetkov}, {Turatto}, \& {Wilke}}]{Hunter09}
{Hunter}, D.~J., {Valenti}, S., {Kotak}, R., {et~al.} 2009, \aap, 508, 371

\bibitem[{{Hurst} {et~al.}(1999){Hurst}, {Armstrong}, \& {Boles}}]{Hurst99}
{Hurst}, G.~M., {Armstrong}, M., \& {Boles}, T. 1999, \iaucirc, 7282

\bibitem[{{Inserra} {et~al.}(2012{\natexlab{a}}){Inserra}, {Baron}, \&
  {Turatto}}]{Inserra12}
{Inserra}, C., {Baron}, E., \& {Turatto}, M. 2012{\natexlab{a}}, \mnras, 422,
  1178

\bibitem[{{Inserra} {et~al.}(2012{\natexlab{b}}){Inserra}, {Turatto},
  {Pastorello}, {Pumo}, {Baron}, {Benetti}, {Cappellaro}, {Taubenberger},
  {Bufano}, {Elias-Rosa}, {Zampieri}, {Harutyunyan}, {Moskvitin}, {Nissinen},
  {Stanishev}, {Tsvetkov}, {Hentunen}, {Komarova}, {Pavlyuk}, {Sokolov}, \&
  {Sokolova}}]{Inserra1209bw}
{Inserra}, C., {Turatto}, M., {Pastorello}, A., {et~al.} 2012{\natexlab{b}},
  \mnras, 422, 1122

\bibitem[{{Jack} {et~al.}(2015){Jack}, {Mittag}, {Schr{\"o}der}, {Schmitt},
  {Hempelmann}, {Gonz{\'a}lez-P{\'e}rez}, {Trinidad}, {Rauw}, \& {Cabrera
  Sixto}}]{Jack15}
{Jack}, D., {Mittag}, M., {Schr{\"o}der}, K.-P., {et~al.} 2015, \mnras, 451,
  4104

\bibitem[{{Jha} {et~al.}(2007){Jha}, {Riess}, \& {Kirshner}}]{Jha07}
{Jha}, S., {Riess}, A.~G., \& {Kirshner}, R.~P. 2007, ApJ, 659, 122

\bibitem[{{Kankare} {et~al.}(2012){Kankare}, {Ergon}, {Bufano}, {Spyromilio},
  {Mattila}, {Chugai}, {Lundqvist}, {Pastorello}, {Kotak}, {Benetti},
  {Botticella}, {Cumming}, {Fransson}, {Fraser}, {Leloudas}, {Miluzio},
  {Sollerman}, {Stritzinger}, {Turatto}, \& {Valenti}}]{Kankare12}
{Kankare}, E., {Ergon}, M., {Bufano}, F., {et~al.} 2012, \mnras, 424, 855

\bibitem[{{Kankare} {et~al.}(2014){Kankare}, {Fraser}, {Ryder},
  {Romero-Ca{\~n}izales}, {Mattila}, {Kotak}, {Laursen}, {Monard}, {Salvo}, \&
  {V{\"a}is{\"a}nen}}]{Kankare14}
{Kankare}, E., {Fraser}, M., {Ryder}, S., {et~al.} 2014, \aap, 572, A75

\bibitem[{{Kasliwal} {et~al.}(2008){Kasliwal}, {Ofek}, {Gal-Yam}, {Rau},
  {Brown}, {Cenko}, {Cameron}, {Quimby}, {Kulkarni}, {Bildsten}, {Milne}, \&
  {Bryngelson}}]{Kasliwal08}
{Kasliwal}, M.~M., {Ofek}, E.~O., {Gal-Yam}, A., {et~al.} 2008, ApJL, 683, L29

\bibitem[{{Kawabata} {et~al.}(2014){Kawabata}, {Akitaya}, {Yamanaka}, {Itoh},
  {Maeda}, {Moritani}, {Ui}, {Kawabata}, {Mori}, {Nogami}, {Nomoto}, {Suzuki},
  {Takaki}, {Tanaka}, {Ueno}, {Chiyonobu}, {Harao}, {Matsui}, {Miyamoto},
  {Nagae}, {Nakashima}, {Nakaya}, {Ohashi}, {Ohsugi}, {Komatsu}, {Sakimoto},
  {Sasada}, {Sato}, {Tanaka}, {Urano}, {Yamashita}, {Yoshida}, {Arai},
  {Ebisuda}, {Fukazawa}, {Fukui}, {Hashimoto}, {Honda}, {Izumiura}, {Kanda},
  {Kawaguchi}, {Kawai}, {Kuroda}, {Masumoto}, {Matsumoto}, {Nakaoka}, {Takata},
  {Uemura}, \& {Yanagisawa}}]{Kawabata14}
{Kawabata}, K.~S., {Akitaya}, H., {Yamanaka}, M., {et~al.} 2014, \apjl, 795, L4

\bibitem[{{Khan} {et~al.}(2011){Khan}, {Prieto}, {Pojma{\'n}ski}, {Stanek},
  {Beacom}, {Szczygiel}, {Pilecki}, {Mogren}, {Eastman}, {Martini}, \&
  {Stoll}}]{Khan11}
{Khan}, R., {Prieto}, J.~L., {Pojma{\'n}ski}, G., {et~al.} 2011, ApJ, 726, 106

\bibitem[{{Kinne}(2012)}]{Kinne12}
{Kinne}, R.~C.~S. 2012, Journal of the American Association of Variable Star
  Observers (JAAVSO), 40, 208

\bibitem[{{Kirshner} {et~al.}(1987){Kirshner}, {Sonneborn}, {Crenshaw}, \&
  {Nassiopoulos}}]{Kirshner87}
{Kirshner}, R.~P., {Sonneborn}, G., {Crenshaw}, D.~M., \& {Nassiopoulos}, G.~E.
  1987, \apj, 320, 602

\bibitem[{{Klein} \& {Chevalier}(1978)}]{Klein78}
{Klein}, R.~I., \& {Chevalier}, R.~A. 1978, \apjl, 223, L109

\bibitem[{{Kotak} {et~al.}(2009){Kotak}, {Meikle}, {Farrah}, {Gerardy},
  {Foley}, {Van Dyk}, {Fransson}, {Lundqvist}, {Sollerman}, {Fesen},
  {Filippenko}, {Mattila}, {Silverman}, {Andersen}, {H{\"o}flich}, {Pozzo}, \&
  {Wheeler}}]{Kotak09}
{Kotak}, R., {Meikle}, W.~P.~S., {Farrah}, D., {et~al.} 2009, \apj, 704, 306

\bibitem[{{Krause} {et~al.}(2008){Krause}, {Tanaka}, {Usuda}, {Hattori},
  {Goto}, {Birkmann}, \& {Nomoto}}]{Krause:2008a}
{Krause}, O., {Tanaka}, M., {Usuda}, T., {et~al.} 2008, \nat, 456, 617

\bibitem[{{Kromer} {et~al.}(2016){Kromer}, {Fremling}, {Pakmor},
  {Taubenberger}, {Amanullah}, {Cenko}, {Fransson}, {Goobar}, {Leloudas},
  {Taddia}, {Roepke}, {Seitenzahl}, {Sim}, \& {Sollerman}}]{Kromer16}
{Kromer}, M., {Fremling}, C., {Pakmor}, R., {et~al.} 2016, ArXiv e-prints,
  arXiv:1604.05730

\bibitem[{{Lennarz} {et~al.}(2012){Lennarz}, {Altmann}, \&
  {Wiebusch}}]{Lennarz12}
{Lennarz}, D., {Altmann}, D., \& {Wiebusch}, C. 2012, \aap, 538, A120

\bibitem[{{Li} {et~al.}(2011){Li}, {Leaman}, {Chornock}, {Filippenko},
  {Poznanski}, {Ganeshalingam}, {Wang}, {Modjaz}, {Jha}, {Foley}, \&
  {Smith}}]{WLi11b}
{Li}, W., {Leaman}, J., {Chornock}, R., {et~al.} 2011, MNRAS, 412, 1441

\bibitem[{{Liu} {et~al.}(2015{\natexlab{a}}){Liu}, {Zhao}, {Huang}, {Wang},
  {Zhang}, {Chen}, \& {Zhang}}]{Liu1512ap}
{Liu}, Z., {Zhao}, X.-L., {Huang}, F., {et~al.} 2015{\natexlab{a}}, Research in
  Astronomy and Astrophysics, 15, 225

\bibitem[{{Liu} {et~al.}(2015{\natexlab{b}}){Liu}, {Zhang}, {Ciabattari},
  {Tomasella}, {Wang}, {Zhao}, {Zhang}, {Xin}, {Wang}, \& {Chang}}]{Liu1513en}
{Liu}, Z.-W., {Zhang}, J.-J., {Ciabattari}, F., {et~al.} 2015{\natexlab{b}},
  \mnras, 452, 838

\bibitem[{{Lunnan} {et~al.}(2013){Lunnan}, {Chornock}, {Berger},
  {Milisavljevic}, {Drout}, {Sanders}, {Challis}, {Czekala}, {Foley}, {Fong},
  {Huber}, {Kirshner}, {Leibler}, {Marion}, {McCrum}, {Narayan}, {Rest},
  {Roth}, {Scolnic}, {Smartt}, {Smith}, {Soderberg}, {Stubbs}, {Tonry},
  {Burgett}, {Chambers}, {Kudritzki}, {Magnier}, \& {Price}}]{Lunnan13}
{Lunnan}, R., {Chornock}, R., {Berger}, E., {et~al.} 2013, \apj, 771, 97

\bibitem[{{Maeda} {et~al.}(2009){Maeda}, {Kawabata}, {Li}, {Tanaka}, {Mazzali},
  {Hattori}, {Nomoto}, \& {Filippenko}}]{Maeda09}
{Maeda}, K., {Kawabata}, K., {Li}, W., {et~al.} 2009, ApJ, 690, 1745

\bibitem[{{Maeda} {et~al.}(2015){Maeda}, {Hattori}, {Milisavljevic},
  {Folatelli}, {Drout}, {Kuncarayakti}, {Margutti}, {Kamble}, {Soderberg},
  {Tanaka}, {Kawabata}, {Kawabata}, {Yamanaka}, {Nomoto}, {Kim}, {Simon},
  {Phillips}, {Parrent}, {Nakaoka}, {Moriya}, {Suzuki}, {Takaki}, {Ishigaki},
  {Sakon}, {Tajitsu}, \& {Iye}}]{Maeda15}
{Maeda}, K., {Hattori}, T., {Milisavljevic}, D., {et~al.} 2015, \apj, 807, 35

\bibitem[{{Magee} {et~al.}(2016){Magee}, {Kotak}, {Sim}, {Kromer},
  {Rabinowitz}, {Smartt}, {Baltay}, {Campbell}, {Chen}, {Fink}, {Gal-Yam},
  {Galbany}, {Hillebrandt}, {Inserra}, {Kankare}, {Le Guillou}, {Lyman},
  {Maguire}, {Pakmor}, {R{\"o}pke}, {Ruiter}, {Seitenzahl}, {Sullivan},
  {Valenti}, \& {Young}}]{Magee16}
{Magee}, M.~R., {Kotak}, R., {Sim}, S.~A., {et~al.} 2016, \aap, 589, A89

\bibitem[{{Maggi} {et~al.}(2016){Maggi}, {Haberl}, {Kavanagh}, {Sasaki},
  {Bozzetto}, {Filipovi{\'c}}, {Vasilopoulos}, {Pietsch}, {Points}, {Chu},
  {Dickel}, {Ehle}, {Williams}, \& {Greiner}}]{Maggi:2016a}
{Maggi}, P., {Haberl}, F., {Kavanagh}, P.~J., {et~al.} 2016, \aap, 585, A162

\bibitem[{{Maguire} {et~al.}(2012){Maguire}, {Sullivan}, {Ellis}, {Nugent},
  {Howell}, {Gal-Yam}, {Cooke}, {Mazzali}, {Pan}, {Dilday}, {Thomas}, {Arcavi},
  {Ben-Ami}, {Bersier}, {Bianco}, {Fulton}, {Hook}, {Horesh}, {Hsiao}, {James},
  {Podsiadlowski}, {Walker}, {Yaron}, {Kasliwal}, {Laher}, {Law}, {Ofek},
  {Poznanski}, \& {Surace}}]{Maguire12}
{Maguire}, K., {Sullivan}, M., {Ellis}, R.~S., {et~al.} 2012, MNRAS, 426, 2359

\bibitem[{{Maguire} {et~al.}(2014){Maguire}, {Sullivan}, {Pan}, {Gal-Yam},
  {Hook}, {Howell}, {Nugent}, {Mazzali}, {Chotard}, {Clubb}, {Filippenko},
  {Kasliwal}, {Kandrashoff}, {Poznanski}, {Saunders}, {Silverman}, {Walker}, \&
  {Xu}}]{Maguire14}
{Maguire}, K., {Sullivan}, M., {Pan}, Y.-C., {et~al.} 2014, MNRAS, 444, 3258

\bibitem[{{Maoz} {et~al.}(2012){Maoz}, {Mannucci}, \& {Brandt}}]{Maoz12}
{Maoz}, D., {Mannucci}, F., \& {Brandt}, T.~D. 2012, MNRAS, 426, 3282

\bibitem[{{Marion} {et~al.}(2013){Marion}, {Vinko}, {Wheeler}, {Foley},
  {Hsiao}, {Brown}, {Challis}, {Filippenko}, {Garnavich}, {Kirshner},
  {Landsman}, {Parrent}, {Pritchard}, {Roming}, {Silverman}, \&
  {Wang}}]{Marion13}
{Marion}, G.~H., {Vinko}, J., {Wheeler}, J.~C., {et~al.} 2013, ApJ, 777, 40

\bibitem[{{Marion} {et~al.}(2014){Marion}, {Vinko}, {Kirshner}, {Foley},
  {Berlind}, {Bieryla}, {Bloom}, {Calkins}, {Challis}, {Chevalier}, {Chornock},
  {Culliton}, {Curtis}, {Esquerdo}, {Everett}, {Falco}, {France}, {Fransson},
  {Friedman}, {Garnavich}, {Leibundgut}, {Meyer}, {Smith}, {Soderberg},
  {Sollerman}, {Starr}, {Szklenar}, {Takats}, \& {Wheeler}}]{Marion14IIb}
{Marion}, G.~H., {Vinko}, J., {Kirshner}, R.~P., {et~al.} 2014, ApJ, 781, 69

\bibitem[{{Marion} {et~al.}(2015){Marion}, {Sand}, {Hsiao}, {Banerjee},
  {Valenti}, {Stritzinger}, {Vink{\'o}}, {Joshi}, {Venkataraman}, {Ashok},
  {Amanullah}, {Binzel}, {Bochanski}, {Bryngelson}, {Burns}, {Drozdov},
  {Fieber-Beyer}, {Graham}, {Howell}, {Johansson}, {Kirshner}, {Milne},
  {Parrent}, {Silverman}, {Vervack}, \& {Wheeler}}]{Marion15}
{Marion}, G.~H., {Sand}, D.~J., {Hsiao}, E.~Y., {et~al.} 2015, ApJ, 798, 39

\bibitem[{{Matheson} {et~al.}(2000{\natexlab{a}}){Matheson}, {Filippenko},
  {Ho}, {Barth}, \& {Leonard}}]{Matheson00b}
{Matheson}, T., {Filippenko}, A.~V., {Ho}, L.~C., {Barth}, A.~J., \& {Leonard},
  D.~C. 2000{\natexlab{a}}, \aj, 120, 1499

\bibitem[{{Matheson} {et~al.}(2000{\natexlab{b}}){Matheson}, {Filippenko},
  {Barth}, {Ho}, {Leonard}, {Bershady}, {Davis}, {Finley}, {Fisher},
  {Gonz{\'a}lez}, {Hawley}, {Koo}, {Li}, {Lonsdale}, {Schlegel}, {Smith},
  {Spinrad}, \& {Wirth}}]{Matheson00a}
{Matheson}, T., {Filippenko}, A.~V., {Barth}, A.~J., {et~al.}
  2000{\natexlab{b}}, \aj, 120, 1487

\bibitem[{{Matheson} {et~al.}(2012){Matheson}, {Joyce}, {Allen}, {Saha},
  {Silva}, {Wood-Vasey}, {Adams}, {Anderson}, {Beck}, {Bentz}, {Bershady},
  {Binkert}, {Butler}, {Camarata}, {Eigenbrot}, {Everett}, {Gallagher},
  {Garnavich}, {Glikman}, {Harbeck}, {Hargis}, {Herbst}, {Horch}, {Howell},
  {Jha}, {Kaczmarek}, {Knezek}, {Manne-Nicholas}, {Mathieu}, {Meixner},
  {Milliman}, {Power}, {Rajagopal}, {Reetz}, {Rhode}, {Schechtman-Rook},
  {Schwamb}, {Schweiker}, {Simmons}, {Simon}, {Summers}, {Young}, {Weyant},
  {Wilcots}, {Will}, \& {Williams}}]{Matheson12}
{Matheson}, T., {Joyce}, R.~R., {Allen}, L.~E., {et~al.} 2012, ApJ, 754, 19

\bibitem[{{Mattila} {et~al.}(2005){Mattila}, {Lundqvist}, {Sollerman}, {Kozma},
  {Baron}, {Fransson}, {Leibundgut}, \& {Nomoto}}]{Mattila05}
{Mattila}, S., {Lundqvist}, P., {Sollerman}, J., {et~al.} 2005, A\&A, 443, 649

\bibitem[{{Mauerhan} {et~al.}(2013){Mauerhan}, {Smith}, {Silverman},
  {Filippenko}, {Morgan}, {Cenko}, {Ganeshalingam}, {Clubb}, {Bloom},
  {Matheson}, \& {Milne}}]{Mauerhan13}
{Mauerhan}, J.~C., {Smith}, N., {Silverman}, J.~M., {et~al.} 2013, \mnras, 431,
  2599

\bibitem[{{Mazzali} {et~al.}(2010){Mazzali}, {Maurer}, {Valenti}, {Kotak}, \&
  {Hunter}}]{Mazzali10}
{Mazzali}, P.~A., {Maurer}, I., {Valenti}, S., {Kotak}, R., \& {Hunter}, D.
  2010, \mnras, 408, 87

\bibitem[{{Mazzali} {et~al.}(2014){Mazzali}, {Sullivan}, {Hachinger}, {Ellis},
  {Nugent}, {Howell}, {Gal-Yam}, {Maguire}, {Cooke}, {Thomas}, {Nomoto}, \&
  {Walker}}]{Mazzali14}
{Mazzali}, P.~A., {Sullivan}, M., {Hachinger}, S., {et~al.} 2014, MNRAS, 439,
  1959

\bibitem[{{Mazzali} {et~al.}(2015){Mazzali}, {Sullivan}, {Filippenko},
  {Garnavich}, {Clubb}, {Maguire}, {Pan}, {Shappee}, {Silverman}, {Benetti},
  {Hachinger}, {Nomoto}, \& {Pian}}]{Mazzali15nebular11fe}
{Mazzali}, P.~A., {Sullivan}, M., {Filippenko}, A.~V., {et~al.} 2015,
  arXiv:1504.04857, arXiv:1504.04857

\bibitem[{{McClelland} {et~al.}(2013){McClelland}, {Garnavich}, {Milne},
  {Shappee}, \& {Pogge}}]{McClelland13}
{McClelland}, C.~M., {Garnavich}, P.~M., {Milne}, P.~A., {Shappee}, B.~J., \&
  {Pogge}, R.~W. 2013, \apj, 767, 119

\bibitem[{{McClelland} {et~al.}(2010){McClelland}, {Garnavich}, {Galbany},
  {Miquel}, {Foley}, {Filippenko}, {Bassett}, {Wheeler}, {Goobar}, {Jha},
  {Sako}, {Frieman}, {Sollerman}, {Vinko}, \& {Schneider}}]{McClelland10}
{McClelland}, C.~M., {Garnavich}, P.~M., {Galbany}, L., {et~al.} 2010, ApJ,
  720, 704

\bibitem[{{McCrum} {et~al.}(2014){McCrum}, {Smartt}, {Kotak}, {Rest},
  {Jerkstrand}, {Inserra}, {Rodney}, {Chen}, {Howell}, {Huber}, {Pastorello},
  {Tonry}, {Bresolin}, {Kudritzki}, {Chornock}, {Berger}, {Smith},
  {Botticella}, {Foley}, {Fraser}, {Milisavljevic}, {Nicholl}, {Riess},
  {Stubbs}, {Valenti}, {Wood-Vasey}, {Wright}, {Young}, {Drout}, {Czekala},
  {Burgett}, {Chambers}, {Draper}, {Flewelling}, {Hodapp}, {Kaiser}, {Magnier},
  {Metcalfe}, {Price}, {Sweeney}, \& {Wainscoat}}]{McCrum14}
{McCrum}, M., {Smartt}, S.~J., {Kotak}, R., {et~al.} 2014, \mnras, 437, 656

\bibitem[{{McCrum} {et~al.}(2015){McCrum}, {Smartt}, {Rest}, {Smith}, {Kotak},
  {Rodney}, {Young}, {Chornock}, {Berger}, {Foley}, {Fraser}, {Wright},
  {Scolnic}, {Tonry}, {Urata}, {Huang}, {Pastorello}, {Botticella}, {Valenti},
  {Mattila}, {Kankare}, {Farrow}, {Huber}, {Stubbs}, {Kirshner}, {Bresolin},
  {Burgett}, {Chambers}, {Draper}, {Flewelling}, {Jedicke}, {Kaiser},
  {Magnier}, {Metcalfe}, {Morgan}, {Price}, {Sweeney}, {Wainscoat}, \&
  {Waters}}]{McCrum15}
{McCrum}, M., {Smartt}, S.~J., {Rest}, A., {et~al.} 2015, \mnras, 448, 1206

\bibitem[{{McCully} {et~al.}(2014){McCully}, {Jha}, {Foley}, {Chornock},
  {Holtzman}, {Balam}, {Branch}, {Filippenko}, {Frieman}, {Fynbo}, {Galbany},
  {Ganeshalingam}, {Garnavich}, {Graham}, {Hsiao}, {Leloudas}, {Leonard}, {Li},
  {Riess}, {Sako}, {Schneider}, {Silverman}, {Sollerman}, {Steele}, {Thomas},
  {Wheeler}, \& {Zheng}}]{McCully14a}
{McCully}, C., {Jha}, S.~W., {Foley}, R.~J., {et~al.} 2014, ApJ, 786, 134

\bibitem[{{Meikle} {et~al.}(1996){Meikle}, {Cumming}, {Geballe}, {Lewis},
  {Walton}, {Balcells}, {Cimatti}, {Croom}, {Dhillon}, {Economou}, {Jenkins},
  {Knapen}, {Meadows}, {Morris}, {Perez-Fournon}, {Shanks}, {Smith}, {Tanvir},
  {Veilleux}, {Vilchez}, {Wall}, \& {Lucey}}]{Meikle96}
{Meikle}, W.~P.~S., {Cumming}, R.~J., {Geballe}, T.~R., {et~al.} 1996, MNRAS,
  281, 263

\bibitem[{{Melandri} {et~al.}(2014){Melandri}, {Pian}, {D'Elia}, {D'Avanzo},
  {Della Valle}, {Mazzali}, {Tagliaferri}, {Cano}, {Levan}, {M{$\Delta$}oller},
  {Amati}, {Bernardini}, {Bersier}, {Bufano}, {Campana}, {Castro-Tirado},
  {Covino}, {Ghirlanda}, {Hurley}, {Malesani}, {Masetti}, {Palazzi},
  {Piranomonte}, {Rossi}, {Salvaterra}, {Starling}, {Tanaka}, {Tanvir}, \&
  {Vergani}}]{Melandri14}
{Melandri}, A., {Pian}, E., {D'Elia}, V., {et~al.} 2014, \aap, 567, A29

\bibitem[{{Milisavljevic} {et~al.}(2013){Milisavljevic}, {Soderberg},
  {Margutti}, {Drout}, {Howie Marion}, {Sanders}, {Hsiao}, {Lunnan},
  {Chornock}, {Fesen}, {Parrent}, {Levesque}, {Berger}, {Foley}, {Challis},
  {Kirshner}, {Dittmann}, {Bieryla}, {Kamble}, {Chakraborti}, {De Rosa},
  {Fausnaugh}, {Hainline}, {Chen}, {Hickox}, {Morrell}, {Phillips}, \&
  {Stritzinger}}]{Danmil13}
{Milisavljevic}, D., {Soderberg}, A.~M., {Margutti}, R., {et~al.} 2013, ApJL,
  770, L38

\bibitem[{{Misra} {et~al.}(2007){Misra}, {Pooley}, {Chandra}, {Bhattacharya},
  {Ray}, {Sagar}, \& {Lewin}}]{Misra07}
{Misra}, K., {Pooley}, D., {Chandra}, P., {et~al.} 2007, \mnras, 381, 280

\bibitem[{{Misra} {et~al.}(2008){Misra}, {Sahu}, {Anupama}, \&
  {Pandey}}]{Misra08}
{Misra}, K., {Sahu}, D.~K., {Anupama}, G.~C., \& {Pandey}, K. 2008, \mnras,
  389, 706

\bibitem[{{Modjaz} {et~al.}(2014){Modjaz}, {Blondin}, {Kirshner}, {Matheson},
  {Berlind}, {Bianco}, {Calkins}, {Challis}, {Garnavich}, {Hicken}, {Jha},
  {Liu}, \& {Marion}}]{Modjaz14}
{Modjaz}, M., {Blondin}, S., {Kirshner}, R.~P., {et~al.} 2014, AJ, 147, 99

\bibitem[{{Morales-Garoffolo} {et~al.}(2014){Morales-Garoffolo}, {Elias-Rosa},
  {Benetti}, {Taubenberger}, {Cappellaro}, {Pastorello}, {Klauser}, {Valenti},
  {Howerton}, {Ochner}, {Schramm}, {Siviero}, {Tartaglia}, \&
  {Tomasella}}]{Morales14}
{Morales-Garoffolo}, A., {Elias-Rosa}, N., {Benetti}, S., {et~al.} 2014,
  \mnras, 445, 1647

\bibitem[{{Munari} {et~al.}(1998){Munari}, {Barbon}, {Piemonte}, {Tomasella},
  \& {Rejkuba}}]{Munari98}
{Munari}, U., {Barbon}, R., {Piemonte}, A., {Tomasella}, L., \& {Rejkuba}, M.
  1998, \aap, 333, 159

\bibitem[{{Munari} {et~al.}(2013){Munari}, {Henden}, {Belligoli}, {Castellani},
  {Cherini}, {Righetti}, \& {Vagnozzi}}]{Munari13}
{Munari}, U., {Henden}, A., {Belligoli}, R., {et~al.} 2013, na, 20, 30

\bibitem[{{Narayan} {et~al.}(2011){Narayan}, {Foley}, {Berger}, {Botticella},
  {Chornock}, {Huber}, {Rest}, {Scolnic}, {Smartt}, {Valenti}, {Soderberg},
  {Burgett}, {Chambers}, {Flewelling}, {Gates}, {Grav}, {Kaiser}, {Kirshner},
  {Magnier}, {Morgan}, {Price}, {Riess}, {Stubbs}, {Sweeney}, {Tonry},
  {Wainscoat}, {Waters}, \& {Wood-Vasey}}]{Narayan11}
{Narayan}, G., {Foley}, R.~J., {Berger}, E., {et~al.} 2011, ApJL, 731, L11

\bibitem[{{Neuhaeuser} {et~al.}(2016){Neuhaeuser}, {Ehrig-Eggert}, \&
  {Kunitzsch}}]{Neuhaeuser:2016a}
{Neuhaeuser}, R., {Ehrig-Eggert}, C., \& {Kunitzsch}, P. 2016, ArXiv e-prints,
  arXiv:1604.03798

\bibitem[{{Nicholl} {et~al.}(2014){Nicholl}, {Smartt}, {Jerkstrand}, {Inserra},
  {Anderson}, {Baltay}, {Benetti}, {Chen}, {Elias-Rosa}, {Feindt}, {Fraser},
  {Gal-Yam}, {Hadjiyska}, {Howell}, {Kotak}, {Lawrence}, {Leloudas},
  {Margheim}, {Mattila}, {McCrum}, {McKinnon}, {Mead}, {Nugent}, {Rabinowitz},
  {Rest}, {Smith}, {Sollerman}, {Sullivan}, {Taddia}, {Valenti}, {Walker}, \&
  {Young}}]{Nicholl14}
{Nicholl}, M., {Smartt}, S.~J., {Jerkstrand}, A., {et~al.} 2014, \mnras, 444,
  2096

\bibitem[{{Nugent} {et~al.}(2011){Nugent}, {Sullivan}, {Cenko}, {Thomas},
  {Kasen}, {Howell}, {Bersier}, {Bloom}, {Kulkarni}, {Kandrashoff},
  {Filippenko}, {Silverman}, {Marcy}, {Howard}, {Isaacson}, {Maguire},
  {Suzuki}, {Tarlton}, {Pan}, {Bildsten}, {Fulton}, {Parrent}, {Sand},
  {Podsiadlowski}, {Bianco}, {Dilday}, {Graham}, {Lyman}, {James}, {Kasliwal},
  {Law}, {Quimby}, {Hook}, {Walker}, {Mazzali}, {Pian}, {Ofek}, {Gal-Yam}, \&
  {Poznanski}}]{Nugent11}
{Nugent}, P.~E., {Sullivan}, M., {Cenko}, S.~B., {et~al.} 2011, Nature, 480,
  344

\bibitem[{{Ofek} {et~al.}(2014){Ofek}, {Zoglauer}, {Boggs}, {Barri{\'e}re},
  {Reynolds}, {Fryer}, {Harrison}, {Cenko}, {Kulkarni}, {Gal-Yam}, {Arcavi},
  {Bellm}, {Bloom}, {Christensen}, {Craig}, {Even}, {Filippenko},
  {Grefenstette}, {Hailey}, {Laher}, {Madsen}, {Nakar}, {Nugent}, {Stern},
  {Sullivan}, {Surace}, \& {Zhang}}]{Ofek14}
{Ofek}, E.~O., {Zoglauer}, A., {Boggs}, S.~E., {et~al.} 2014, \apj, 781, 42

\bibitem[{{Parrent} {et~al.}(2016{\natexlab{a}}){Parrent}, {Milisavljevic},
  {Soderberg}, \& {Parthasarathy}}]{Parrent16}
{Parrent}, J.~T., {Milisavljevic}, D., {Soderberg}, A.~M., \& {Parthasarathy},
  M. 2016{\natexlab{a}}, \apj, 820, 75

\bibitem[{{Parrent} {et~al.}(2012){Parrent}, {Howell}, {Friesen}, {Thomas},
  {Fesen}, {Milisavljevic}, {Bianco}, {Dilday}, {Nugent}, {Baron}, {Arcavi},
  {Ben-Ami}, {Bersier}, {Bildsten}, {Bloom}, {Cao}, {Cenko}, {Filippenko},
  {Gal-Yam}, {Kasliwal}, {Konidaris}, {Kulkarni}, {Law}, {Levitan}, {Maguire},
  {Mazzali}, {Ofek}, {Pan}, {Polishook}, {Poznanski}, {Quimby}, {Silverman},
  {Sternberg}, {Sullivan}, {Walker}, {Xu}, {Buton}, \& {Pereira}}]{Parrent12}
{Parrent}, J.~T., {Howell}, D.~A., {Friesen}, B., {et~al.} 2012, ApJL, 752, L26

\bibitem[{{Parrent} {et~al.}(2016{\natexlab{b}}){Parrent}, {Howell}, {Fesen},
  {Parker}, {Bianco}, {Dilday}, {Sand}, {Valenti}, {Vink{\'o}}, {Berlind},
  {Challis}, {Milisavljevic}, {Sanders}, {Marion}, {Wheeler}, {Brown},
  {Calkins}, {Friesen}, {Kirshner}, {Pritchard}, {Quimby}, \&
  {Roming}}]{Parrent1612dn}
{Parrent}, J.~T., {Howell}, D.~A., {Fesen}, R.~A., {et~al.} 2016{\natexlab{b}},
  \mnras, 457, 3702

\bibitem[{{Pereira} {et~al.}(2013){Pereira}, {Thomas}, {Aldering}, {Antilogus},
  {Baltay}, {Benitez-Herrera}, {Bongard}, {Buton}, {Canto}, {Cellier-Holzem},
  {Chen}, {Childress}, {Chotard}, {Copin}, {Fakhouri}, {Fink}, {Fouchez},
  {Gangler}, {Guy}, {Hillebrandt}, {Hsiao}, {Kerschhaggl}, {Kowalski},
  {Kromer}, {Nordin}, {Nugent}, {Paech}, {Pain}, {P{\'e}contal}, {Perlmutter},
  {Rabinowitz}, {Rigault}, {Runge}, {Saunders}, {Smadja}, {Tao},
  {Taubenberger}, {Tilquin}, \& {Wu}}]{Pereira13}
{Pereira}, R., {Thomas}, R.~C., {Aldering}, G., {et~al.} 2013, A\&A, 554, A27

\bibitem[{{Pritchard} {et~al.}(2012){Pritchard}, {Roming}, {Brown}, {Kuin},
  {Bayless}, {Holland}, {Immler}, {Milne}, \& {Oates}}]{Pritchard12}
{Pritchard}, T.~A., {Roming}, P.~W.~A., {Brown}, P.~J., {et~al.} 2012, \apj,
  750, 128

\bibitem[{{Qiu} {et~al.}(1999){Qiu}, {Li}, {Qiao}, \& {Hu}}]{Qiu99}
{Qiu}, Y., {Li}, W., {Qiao}, Q., \& {Hu}, J. 1999, \aj, 117, 736

\bibitem[{{Rau} {et~al.}(2009){Rau}, {Kulkarni}, {Law}, {Bloom}, {Ciardi},
  {Djorgovski}, {Fox}, {Gal-Yam}, {Grillmair}, {Kasliwal}, {Nugent}, {Ofek},
  {Quimby}, {Reach}, {Shara}, {Bildsten}, {Cenko}, {Drake}, {Filippenko},
  {Helfand}, {Helou}, {Howell}, {Poznanski}, \& {Sullivan}}]{Rau09}
{Rau}, A., {Kulkarni}, S.~R., {Law}, N.~M., {et~al.} 2009, PASP, 121, 1334

\bibitem[{{Rein}(2012)}]{Rein12}
{Rein}, H. 2012, ArXiv e-prints, arXiv:1211.7121

\bibitem[{{Rest} {et~al.}(2011){Rest}, {Foley}, {Gezari}, {Narayan}, {Draine},
  {Olsen}, {Huber}, {Matheson}, {Garg}, {Welch}, {Becker}, {Challis},
  {Clocchiatti}, {Cook}, {Damke}, {Meixner}, {Miknaitis}, {Minniti}, {Morelli},
  {Nikolaev}, {Pignata}, {Prieto}, {Smith}, {Stubbs}, {Suntzeff}, {Walker},
  {Wood-Vasey}, {Zenteno}, {Wyrzykowski}, {Udalski}, {Szyma{\'n}ski}, {Kubiak},
  {Pietrzy{\'n}ski}, {Soszy{\'n}ski}, {Szewczyk}, {Ulaczyk}, \&
  {Poleski}}]{Rest11}
{Rest}, A., {Foley}, R.~J., {Gezari}, S., {et~al.} 2011, \apj, 729, 88

\bibitem[{{Rest} {et~al.}(2014){Rest}, {Scolnic}, {Foley}, {Huber}, {Chornock},
  {Narayan}, {Tonry}, {Berger}, {Soderberg}, {Stubbs}, {Riess}, {Kirshner},
  {Smartt}, {Schlafly}, {Rodney}, {Botticella}, {Brout}, {Challis}, {Czekala},
  {Drout}, {Hudson}, {Kotak}, {Leibler}, {Lunnan}, {Marion}, {McCrum},
  {Milisavljevic}, {Pastorello}, {Sanders}, {Smith}, {Stafford}, {Thilker},
  {Valenti}, {Wood-Vasey}, {Zheng}, {Burgett}, {Chambers}, {Denneau}, {Draper},
  {Flewelling}, {Hodapp}, {Kaiser}, {Kudritzki}, {Magnier}, {Metcalfe},
  {Price}, {Sweeney}, {Wainscoat}, \& {Waters}}]{Rest14}
{Rest}, A., {Scolnic}, D., {Foley}, R.~J., {et~al.} 2014, \apj, 795, 44

\bibitem[{{Richardson} {et~al.}(2001){Richardson}, {Thomas}, {Casebeer},
  {Blankenship}, {Ratowt}, {Baron}, \& {Branch}}]{Richardson01}
{Richardson}, D., {Thomas}, R.~C., {Casebeer}, D., {et~al.} 2001, in Bulletin
  of the American Astronomical Society, Vol.~33, American Astronomical Society
  Meeting Abstracts, 1428

\bibitem[{{Richmond} \& {Smith}(2012)}]{Richmond12}
{Richmond}, M.~W., \& {Smith}, H.~A. 2012, Journal of the American Association
  of Variable Star Observers (JAAVSO), 40, 872

\bibitem[{{Rigon} {et~al.}(2003){Rigon}, {Turatto}, {Benetti}, {Pastorello},
  {Cappellaro}, {Aretxaga}, {Vega}, {Chavushyan}, {Patat}, {Danziger}, \&
  {Salvo}}]{Rigon03}
{Rigon}, L., {Turatto}, M., {Benetti}, S., {et~al.} 2003, \mnras, 340, 191

\bibitem[{{Roy} {et~al.}(2011){Roy}, {Kumar}, {Benetti}, {Pastorello}, {Yuan},
  {Brown}, {Immler}, {Fatkhullin}, {Moskvitin}, {Maund}, {Akerlof}, {Wheeler},
  {Sokolov}, {Quimby}, {Bufano}, {Kumar}, {Misra}, {Pandey}, {Elias-Rosa},
  {Roming}, \& {Sagar}}]{Roy11}
{Roy}, R., {Kumar}, B., {Benetti}, S., {et~al.} 2011, \apj, 736, 76

\bibitem[{{Roy} {et~al.}(2013){Roy}, {Kumar}, {Maund}, {Schady}, {Olivares},
  {Malesani}, {Leloudas}, {Nandi}, {Tanvir}, {Milisavljevic}, {Hjorth},
  {Misra}, {Kumar}, {Pandey}, {Sagar}, \& {Chandola}}]{Roy13}
{Roy}, R., {Kumar}, B., {Maund}, J.~R., {et~al.} 2013, \mnras, 434, 2032

\bibitem[{{Rudy} {et~al.}(2002){Rudy}, {Lynch}, {Mazuk}, {Venturini},
  {Puetter}, \& {H{\"o}flich}}]{Rudy02}
{Rudy}, R.~J., {Lynch}, D.~K., {Mazuk}, S., {et~al.} 2002, ApJ, 565, 413

\bibitem[{{Ruiz-Lapuente}(2004)}]{RuizLapuente:2004a}
{Ruiz-Lapuente}, P. 2004, \apj, 612, 357

\bibitem[{{Sadakane} {et~al.}(1996){Sadakane}, {Yokoo}, {Arimoto}, {Matsumoto},
  {Honda}, {Tanabe}, {Wakamatsu}, {Nishida}, {Yoshida}, \&
  {Takada-Hidai}}]{Sadakane96}
{Sadakane}, K., {Yokoo}, T., {Arimoto}, J., {et~al.} 1996, \pasj, 48, 51

\bibitem[{{Sahu} {et~al.}(2013){Sahu}, {Anupama}, \& {Chakradhari}}]{Sahu13}
{Sahu}, D.~K., {Anupama}, G.~C., \& {Chakradhari}, N.~K. 2013, \mnras, 433, 2

\bibitem[{{Sahu} {et~al.}(2011){Sahu}, {Gurugubelli}, {Anupama}, \&
  {Nomoto}}]{Sahu11}
{Sahu}, D.~K., {Gurugubelli}, U.~K., {Anupama}, G.~C., \& {Nomoto}, K. 2011,
  \mnras, 413, 2583

\bibitem[{{Sahu} {et~al.}(2009){Sahu}, {Tanaka}, {Anupama}, {Gurugubelli}, \&
  {Nomoto}}]{Sahu09}
{Sahu}, D.~K., {Tanaka}, M., {Anupama}, G.~C., {Gurugubelli}, U.~K., \&
  {Nomoto}, K. 2009, \apj, 697, 676

\bibitem[{{Sahu} {et~al.}(2008){Sahu}, {Tanaka}, {Anupama}, {Kawabata},
  {Maeda}, {Tominaga}, {Nomoto}, {Mazzali}, \& {Prabhu}}]{Sahu08}
{Sahu}, D.~K., {Tanaka}, M., {Anupama}, G.~C., {et~al.} 2008, ApJ, 680, 580

\bibitem[{{Salamanca}(2000)}]{Salamanca00}
{Salamanca}, I. 2000, \memsai, 71, 317

\bibitem[{{Salamanca} {et~al.}(2002){Salamanca}, {Terlevich}, \&
  {Tenorio-Tagle}}]{Salamanca02}
{Salamanca}, I., {Terlevich}, R.~J., \& {Tenorio-Tagle}, G. 2002, \mnras, 330,
  844

\bibitem[{{Sasdelli} {et~al.}(2016){Sasdelli}, {Ishida}, {Hillebrandt},
  {Ashall}, {Mazzali}, \& {Prentice}}]{Sasdelli16}
{Sasdelli}, M., {Ishida}, E.~E.~O., {Hillebrandt}, W., {et~al.} 2016, ArXiv
  e-prints, arXiv:1604.03899

\bibitem[{{Sasdelli} {et~al.}(2015){Sasdelli}, {Hillebrandt}, {Aldering},
  {Antilogus}, {Aragon}, {Bailey}, {Baltay}, {Benitez-Herrera}, {Bongard},
  {Buton}, {Canto}, {Cellier-Holzem}, {Chen}, {Childress}, {Chotard}, {Copin},
  {Fakhouri}, {Feindt}, {Fink}, {Fleury}, {Fouchez}, {Gangler}, {Guy},
  {Ishida}, {Kim}, {Kowalski}, {Kromer}, {Lombardo}, {Mazzali}, {Nordin},
  {Pain}, {P{\'e}contal}, {Pereira}, {Perlmutter}, {Rabinowitz}, {Rigault},
  {Runge}, {Saunders}, {Scalzo}, {Smadja}, {Suzuki}, {Tao}, {Taubenberger},
  {Thomas}, {Tilquin}, \& {Weaver}}]{Sasdelli15PCA}
{Sasdelli}, M., {Hillebrandt}, W., {Aldering}, G., {et~al.} 2015, MNRAS, 447,
  1247

\bibitem[{{Schawinski} {et~al.}(2008){Schawinski}, {Justham}, {Wolf},
  {Podsiadlowski}, {Sullivan}, {Steenbrugge}, {Bell}, {R{\"o}ser}, {Walker},
  {Astier}, {Balam}, {Balland}, {Carlberg}, {Conley}, {Fouchez}, {Guy},
  {Hardin}, {Hook}, {Howell}, {Pain}, {Perrett}, {Pritchet}, {Regnault}, \&
  {Yi}}]{Schawinski08}
{Schawinski}, K., {Justham}, S., {Wolf}, C., {et~al.} 2008, Science, 321, 223

\bibitem[{{Schulze} {et~al.}(2014){Schulze}, {Malesani}, {Cucchiara}, {Tanvir},
  {Kr{\"u}hler}, {de Ugarte Postigo}, {Leloudas}, {Lyman}, {Bersier},
  {Wiersema}, {Perley}, {Schady}, {Gorosabel}, {Anderson}, {Castro-Tirado},
  {Cenko}, {De Cia}, {Ellerbroek}, {Fynbo}, {Greiner}, {Hjorth}, {Kann},
  {Kaper}, {Klose}, {Levan}, {Mart{\'{\i}}n}, {O'Brien}, {Page}, {Pignata},
  {Rapaport}, {S{\'a}nchez-Ram{\'{\i}}rez}, {Sollerman}, {Smith}, {Sparre},
  {Th{\"o}ne}, {Watson}, {Xu}, {Bauer}, {Bayliss}, {Bj{\"o}rnsson}, {Bremer},
  {Cano}, {Covino}, {D'Elia}, {Frail}, {Geier}, {Goldoni}, {Hartoog},
  {Jakobsson}, {Korhonen}, {Lee}, {Milvang-Jensen}, {Nardini}, {Nicuesa
  Guelbenzu}, {Oguri}, {Pandey}, {Petitpas}, {Rossi}, {Sandberg}, {Schmidl},
  {Tagliaferri}, {Tilanus}, {Winters}, {Wright}, \& {Wuyts}}]{Schulze14}
{Schulze}, S., {Malesani}, D., {Cucchiara}, A., {et~al.} 2014, \aap, 566, A102

\bibitem[{{Seaman} {et~al.}(2011){Seaman}, {Williams}, {Allan}, {Barthelmy},
  {Bloom}, {Brewer}, {Denny}, {Fitzpatrick}, {Graham}, {Gray}, {Hessman},
  {Marka}, {Rots}, {Vestrand}, \& {Wozniak}}]{Seaman:2011a}
{Seaman}, R., {Williams}, R., {Allan}, A., {et~al.} 2011, {Sky Event Reporting
  Metadata Version 2.0}, IVOA Recommendation 11 July 2011, , , arXiv:1110.0523

\bibitem[{{Shappee} {et~al.}(2014){Shappee}, {Prieto}, {Stanek}, {Kochanek},
  {Holoien}, {Jencson}, {Basu}, {Beacom}, {Szczygiel}, {Pojmanski},
  {Brimacombe}, {Dubberley}, {Elphick}, {Foale}, {Hawkins}, {Mullins},
  {Rosing}, {Ross}, \& {Walker}}]{Shappee14}
{Shappee}, B., {Prieto}, J., {Stanek}, K.~Z., {et~al.} 2014, in American
  Astronomical Society Meeting Abstracts, Vol. 223, American Astronomical
  Society Meeting Abstracts \#223, 236.03

\bibitem[{{Shappee} {et~al.}(2013){Shappee}, {Stanek}, {Pogge}, \&
  {Garnavich}}]{Shappee13}
{Shappee}, B.~J., {Stanek}, K.~Z., {Pogge}, R.~W., \& {Garnavich}, P.~M. 2013,
  ApJL, 762, L5

\bibitem[{{Silverman} {et~al.}(2012{\natexlab{a}}){Silverman}, {Kong}, \&
  {Filippenko}}]{Silverman12maxlight}
{Silverman}, J.~M., {Kong}, J.~J., \& {Filippenko}, A.~V. 2012{\natexlab{a}},
  MNRAS, 425, 1819

\bibitem[{{Silverman} {et~al.}(2012{\natexlab{b}}){Silverman}, {Ganeshalingam},
  {Cenko}, {Filippenko}, {Li}, {Barth}, {Carson}, {Childress}, {Clubb},
  {Cucchiara}, {Graham}, {Marion}, {Nguyen}, {Pei}, {Tucker}, {Vinko},
  {Wheeler}, \& {Worseck}}]{Silverman12a}
{Silverman}, J.~M., {Ganeshalingam}, M., {Cenko}, S.~B., {et~al.}
  2012{\natexlab{b}}, ApJL, 756, L7

\bibitem[{{Smartt} {et~al.}(2015){Smartt}, {Valenti}, {Fraser}, {Inserra},
  {Young}, {Sullivan}, {Pastorello}, {Benetti}, {Gal-Yam}, {Knapic},
  {Molinaro}, {Smareglia}, {Smith}, {Taubenberger}, {Yaron}, {Anderson},
  {Ashall}, {Balland}, {Baltay}, {Barbarino}, {Bauer}, {Baumont}, {Bersier},
  {Blagorodnova}, {Bongard}, {Botticella}, {Bufano}, {Bulla}, {Cappellaro},
  {Campbell}, {Cellier-Holzem}, {Chen}, {Childress}, {Clocchiatti},
  {Contreras}, {Dall'Ora}, {Danziger}, {de Jaeger}, {De Cia}, {Della Valle},
  {Dennefeld}, {Elias-Rosa}, {Elman}, {Feindt}, {Fleury}, {Gall},
  {Gonzalez-Gaitan}, {Galbany}, {Morales Garoffolo}, {Greggio}, {Guillou},
  {Hachinger}, {Hadjiyska}, {Hage}, {Hillebrandt}, {Hodgkin}, {Hsiao}, {James},
  {Jerkstrand}, {Kangas}, {Kankare}, {Kotak}, {Kromer}, {Kuncarayakti},
  {Leloudas}, {Lundqvist}, {Lyman}, {Hook}, {Maguire}, {Manulis}, {Margheim},
  {Mattila}, {Maund}, {Mazzali}, {McCrum}, {McKinnon}, {Moreno-Raya},
  {Nicholl}, {Nugent}, {Pain}, {Pignata}, {Phillips}, {Polshaw}, {Pumo},
  {Rabinowitz}, {Reilly}, {Romero-Ca{\~n}izales}, {Scalzo}, {Schmidt},
  {Schulze}, {Sim}, {Sollerman}, {Taddia}, {Tartaglia}, {Terreran},
  {Tomasella}, {Turatto}, {Walker}, {Walton}, {Wyrzykowski}, {Yuan}, \&
  {Zampieri}}]{Smartt15PESSTO}
{Smartt}, S.~J., {Valenti}, S., {Fraser}, M., {et~al.} 2015, \aap, 579, A40

\bibitem[{{Smith} {et~al.}(2015){Smith}, {Mauerhan}, {Cenko}, {Kasliwal},
  {Silverman}, {Filippenko}, {Gal-Yam}, {Clubb}, {Graham}, {Leonard}, {Horst},
  {Williams}, {Andrews}, {Kulkarni}, {Nugent}, {Sullivan}, {Maguire}, {Xu}, \&
  {Ben-Ami}}]{Smith15}
{Smith}, N., {Mauerhan}, J.~C., {Cenko}, S.~B., {et~al.} 2015, \mnras, 449,
  1876

\bibitem[{{Smitka} {et~al.}(2016){Smitka}, {Brown}, {Kuin}, \&
  {Suntzeff}}]{Smitka16}
{Smitka}, M.~T., {Brown}, P.~J., {Kuin}, P., \& {Suntzeff}, N.~B. 2016, \pasp,
  128, 034501

\bibitem[{{Soderberg} {et~al.}(2008{\natexlab{a}}){Soderberg}, {Berger},
  {Page}, {Schady}, {Parrent}, {Pooley}, {Wang}, {Ofek}, {Cucchiara}, {Rau},
  {Waxman}, {Simon}, {Bock}, {Milne}, {Page}, {Barentine}, {Barthelmy},
  {Beardmore}, {Bietenholz}, {Brown}, {Burrows}, {Burrows}, {Byrngelson},
  {Cenko}, {Chandra}, {Cummings}, {Fox}, {Gal-Yam}, {Gehrels}, {Immler},
  {Kasliwal}, {Kong}, {Krimm}, {Kulkarni}, {Maccarone}, {M{\'e}sz{\'a}ros},
  {Nakar}, {O'Brien}, {Overzier}, {de Pasquale}, {Racusin}, {Rea}, \&
  {York}}]{Soderberg08D}
{Soderberg}, A.~M., {Berger}, E., {Page}, K.~L., {et~al.} 2008{\natexlab{a}},
  Nature, 453, 469

\bibitem[{{Soderberg} {et~al.}(2008{\natexlab{b}}){Soderberg}, {Berger},
  {Page}, {Schady}, {Parrent}, {Pooley}, {Wang}, {Ofek}, {Cucchiara}, {Rau},
  {Waxman}, {Simon}, {Bock}, {Milne}, {Page}, {Barentine}, {Barthelmy},
  {Beardmore}, {Bietenholz}, {Brown}, {Burrows}, {Burrows}, {Byrngelson},
  {Cenko}, {Chandra}, {Cummings}, {Fox}, {Gal-Yam}, {Gehrels}, {Immler},
  {Kasliwal}, {Kong}, {Krimm}, {Kulkarni}, {Maccarone}, {M{\'e}sz{\'a}ros},
  {Nakar}, {O'Brien}, {Overzier}, {de Pasquale}, {Racusin}, {Rea}, \&
  {York}}]{Soderberg08}
---. 2008{\natexlab{b}}, Nature, 453, 469

\bibitem[{{Sollerman} {et~al.}(1998){Sollerman}, {Leibundgut}, \&
  {Spyromilio}}]{Sollerman98}
{Sollerman}, J., {Leibundgut}, B., \& {Spyromilio}, J. 1998, \aap, 337, 207

\bibitem[{{Spyromilio} {et~al.}(2004){Spyromilio}, {Gilmozzi}, {Sollerman},
  {Leibundgut}, {Fransson}, \& {Cuby}}]{Spyromilio04}
{Spyromilio}, J., {Gilmozzi}, R., {Sollerman}, J., {et~al.} 2004, \aap, 426,
  547

\bibitem[{{Strolger} {et~al.}(2002){Strolger}, {Smith}, {Suntzeff}, {Phillips},
  {Aldering}, {Nugent}, {Knop}, {Perlmutter}, {Schommer}, {Ho}, {Hamuy},
  {Krisciunas}, {Germany}, {Covarrubias}, {Candia}, {Athey}, {Blanc},
  {Bonacic}, {Bowers}, {Conley}, {Dahl{\'e}n}, {Freedman}, {Galaz}, {Gates},
  {Goldhaber}, {Goobar}, {Groom}, {Hook}, {Marzke}, {Mateo}, {McCarthy},
  {M{\'e}ndez}, {Muena}, {Persson}, {Quimby}, {Roth}, {Ruiz-Lapuente},
  {Seguel}, {Szentgyorgyi}, {von Braun}, {Wood-Vasey}, \& {York}}]{Strolger02}
{Strolger}, L.-G., {Smith}, R.~C., {Suntzeff}, N.~B., {et~al.} 2002, AJ, 124,
  2905

\bibitem[{{Szalai} {et~al.}(2011){Szalai}, {Vink{\'o}}, {Balog},
  {G{\'a}sp{\'a}r}, {Block}, \& {Kiss}}]{Szalai11}
{Szalai}, T., {Vink{\'o}}, J., {Balog}, Z., {et~al.} 2011, \aap, 527, A61

\bibitem[{{Szalai} {et~al.}(2015){Szalai}, {Vink{\'o}}, {S{\'a}rneczky},
  {Tak{\'a}ts}, {Benk{\H o}}, {Kelemen}, {Kuli}, {Silverman}, {Marion}, \&
  {Wheeler}}]{Szalai15}
{Szalai}, T., {Vink{\'o}}, J., {S{\'a}rneczky}, K., {et~al.} 2015, \mnras, 453,
  2103

\bibitem[{{Taddia} {et~al.}(2012){Taddia}, {Stritzinger}, {Sollerman},
  {Phillips}, {Anderson}, {Ergon}, {Folatelli}, {Fransson}, {Freedman},
  {Hamuy}, {Morrell}, {Pastorello}, {Persson}, \& {Gonzalez}}]{Taddia1206}
{Taddia}, F., {Stritzinger}, M.~D., {Sollerman}, J., {et~al.} 2012, \aap, 537,
  A140

\bibitem[{{Takaki} {et~al.}(2013){Takaki}, {Kawabata}, {Yamanaka}, {Maeda},
  {Tanaka}, {Akitaya}, {Fukazawa}, {Itoh}, {Kinugasa}, {Moritani}, {Ohsugi},
  {Sasada}, {Uemura}, {Ueno}, {Ui}, {Urano}, {Yoshida}, \& {Nomoto}}]{Takaki13}
{Takaki}, K., {Kawabata}, K.~S., {Yamanaka}, M., {et~al.} 2013, ApJL, 772, L17

\bibitem[{{Tammann} \& {Reindl}(2011)}]{Tammann11}
{Tammann}, G.~A., \& {Reindl}, B. 2011, ArXiv e-prints, arXiv:1112.0439

\bibitem[{{Taubenberger} {et~al.}(2015){Taubenberger}, {Elias-Rosa},
  {Kerzendorf}, {Hachinger}, {Spyromilio}, {Fransson}, {Kromer}, {Ruiter},
  {Seitenzahl}, {Benetti}, {Cappellaro}, {Pastorello}, {Turatto}, \&
  {Marchetti}}]{Taubenberger15}
{Taubenberger}, S., {Elias-Rosa}, N., {Kerzendorf}, W.~E., {et~al.} 2015,
  \mnras, 448, L48

\bibitem[{{Telesco} {et~al.}(2015){Telesco}, {H{\"o}flich}, {Li},
  {{\'A}lvarez}, {Wright}, {Barnes}, {Fern{\'a}ndez}, {Hough}, {Levenson},
  {Mari{\~n}as}, {Packham}, {Pantin}, {Rebolo}, {Roche}, \&
  {Zhang}}]{Telesco15}
{Telesco}, C.~M., {H{\"o}flich}, P., {Li}, D., {et~al.} 2015, \apj, 798, 93

\bibitem[{{Tsvetkov} {et~al.}(2004){Tsvetkov}, {Pavlyuk}, \&
  {Bartunov}}]{Tsvetkov04}
{Tsvetkov}, D.~Y., {Pavlyuk}, N.~N., \& {Bartunov}, O.~S. 2004, Astronomy
  Letters, 30, 729

\bibitem[{{Vacca} {et~al.}(2015){Vacca}, {Hamilton}, {Savage}, {Shenoy},
  {Becklin}, {McLean}, {Logsdon}, {Marion}, {Ashok}, {Banerjee}, {Evans},
  {Fox}, {Garnavich}, {Gehrz}, {Greenhouse}, {Helton}, {Kirshner}, {Shenoy},
  {Smith}, {Spyromilio}, {Starrfield}, {Wooden}, \& {Woodward}}]{Vacca15}
{Vacca}, W.~D., {Hamilton}, R.~T., {Savage}, M., {et~al.} 2015, \apj, 804, 66

\bibitem[{{Vink{\'o}} {et~al.}(2012){Vink{\'o}}, {S{\'a}rneczky}, {Tak{\'a}ts},
  {Marion}, {Heged{\"u}s}, {B{\'{\i}}r{\'o}}, {Borkovits}, {Szegedi-Elek},
  {Farkas}, {Klagyivik}, {Kiss}, {Kov{\'a}cs}, {P{\'a}l}, {Szak{\'a}ts},
  {Szalai}, {Szalai}, {Szatm{\'a}ry}, {Szing}, {Vida}, \& {Wheeler}}]{Vinko12}
{Vink{\'o}}, J., {S{\'a}rneczky}, K., {Tak{\'a}ts}, K., {et~al.} 2012, A\&A,
  546, A12

\bibitem[{{Wang} {et~al.}(1996){Wang}, {Wheeler}, {Kirshner}, {Challis},
  {Filippenko}, {Fransson}, {Panagia}, {Phillips}, \& {Suntzeff}}]{WangL96}
{Wang}, L., {Wheeler}, J.~C., {Kirshner}, R.~P., {et~al.} 1996, \apj, 466, 998

\bibitem[{{Wang} {et~al.}(2013){Wang}, {Wang}, {Filippenko}, {Zhang}, \&
  {Zhao}}]{WangX13}
{Wang}, X., {Wang}, L., {Filippenko}, A.~V., {Zhang}, T., \& {Zhao}, X. 2013,
  Science, 340, 170

\bibitem[{{Wang} {et~al.}(2009){Wang}, {Filippenko}, {Ganeshalingam}, {Li},
  {Silverman}, {Wang}, {Chornock}, {Foley}, {Gates}, {Macomber}, {Serduke},
  {Steele}, \& {Wong}}]{WangX09Subtype}
{Wang}, X., {Filippenko}, A.~V., {Ganeshalingam}, M., {et~al.} 2009, ApJL, 699,
  L139

\bibitem[{{Wright} {et~al.}(2011){Wright}, {Fakhouri}, {Marcy}, {Han}, {Feng},
  {Johnson}, {Howard}, {Fischer}, {Valenti}, {Anderson}, \&
  {Piskunov}}]{Wright:2011a}
{Wright}, J.~T., {Fakhouri}, O., {Marcy}, G.~W., {et~al.} 2011, \pasp, 123, 412

\bibitem[{{Yamanaka} {et~al.}(2009){Yamanaka}, {Kawabata}, {Kinugasa},
  {Tanaka}, {Imada}, {Maeda}, {Nomoto}, {Arai}, {Chiyonobu}, {Fukazawa},
  {Hashimoto}, {Honda}, {Ikejiri}, {Itoh}, {Kamata}, {Kawai}, {Komatsu},
  {Konishi}, {Kuroda}, {Miyamoto}, {Miyazaki}, {Nagae}, {Nakaya}, {Ohsugi},
  {Omodaka}, {Sakai}, {Sasada}, {Suzuki}, {Taguchi}, {Takahashi}, {Tanaka},
  {Uemura}, {Yamashita}, {Yanagisawa}, \& {Yoshida}}]{Yamanaka09a}
{Yamanaka}, M., {Kawabata}, K.~S., {Kinugasa}, K., {et~al.} 2009, ApJL, 707,
  L118

\bibitem[{{Yaron} \& {Gal-Yam}(2012)}]{WISEREP}
{Yaron}, O., \& {Gal-Yam}, A. 2012, PASP, 124, 668

\bibitem[{{Yuan} {et~al.}(2010){Yuan}, {Quimby}, {Wheeler}, {Vink{\'o}},
  {Chatzopoulos}, {Akerlof}, {Kulkarni}, {Miller}, {McKay}, \&
  {Aharonian}}]{Yuan10}
{Yuan}, F., {Quimby}, R.~M., {Wheeler}, J.~C., {et~al.} 2010, ApJ, 715, 1338

\end{thebibliography}

\capstartfalse
\begin{deluxetable}{lr}
\tabletypesize{\footnotesize}
\tablecolumns{2} 
\tablewidth{0pt} 
\tablecaption{An Incomplete List of Publicly Inaccessible Spectra}
\tablehead{\colhead{Supernova Name} & \colhead{Reference}}
\startdata
1983V\tablenotemark{a} & \citet{Clocchiatti97} \\
1987A\tablenotemark{a} & \citet{WangL96} \\
1991D\tablenotemark{a} & \citet{Benetti02} \\
1991T\tablenotemark{a} & \citet{Meikle96} \\
1992ar & \citet{Clocchiatti00} \\
1994D\tablenotemark{a} & \citet{Meikle96} \\
1995D\tablenotemark{a} & \citet{Sadakane96} \\
1995N & \citet{Fransson02} \\
1995V & \citet{Fassia98} \\
1995al\tablenotemark{a} & \citet{Anupama97} \\
1996N & \citet{Sollerman98} \\
1996cb\tablenotemark{a} & \citet{Qiu99} \\
1997X\tablenotemark{a} & \citet{Munari98} \\
1997Y\tablenotemark{a} & \citet{Anupama97} \\
1997ab\tablenotemark{a} & \citet{Salamanca00} \\
1997bp\tablenotemark{a} & \citet{Anupama97} \\
1997eg & \citet{Salamanca02} \\
1998bu\tablenotemark{a} & \citet{Hernandez00} \\ 
& \citet{Spyromilio04} \\
1999E\tablenotemark{a} & \citet{Rigon03} \\
1999aw & \citet{Strolger02} \\
1999by\tablenotemark{a} & \citet{Hoflich02} \\
1999el & \citet{DiCarlo02} \\
1999em\tablenotemark{a} & \citet{Elmhamdi03} \\
2000cx\tablenotemark{a} & \citet{Rudy02} \\
2000ew\tablenotemark{a} & \citet{Gerardy02} \\
2001el\tablenotemark{a} & \citet{Mattila05} \\
2002fk\tablenotemark{a} & \citet{Cartier14} \\
2003hx\tablenotemark{a} & \citet{Misra08} \\
2003ma & \citet{Rest11} \\
2004dj\tablenotemark{a} & \citet{Szalai11} \\
2004et\tablenotemark{a} & \citet{Misra07} \\
 & \citet{Kotak09} \\
2005am\tablenotemark{a} & \citet{Brown05} \\
2005df\tablenotemark{a} & \citet{Diamond15} \\
2005ke\tablenotemark{a} & \citet{Kankare14} \\
2005hk\tablenotemark{a} & \citet{Sahu08} \\
&  \citet{McCully14a} \\
2006bc & \citet{Gallagher12} \\
2006gz\tablenotemark{a} & \citet{Maeda09} \\
2007ax\tablenotemark{a} & \citet{Kasliwal08} \\
2007gr\tablenotemark{a} & \citet{Hunter09} \\
 & \citet{Mazzali10} \\
 & \citet{Chen14} \\
2007od & \citet{Inserra12} \\
2007pk\tablenotemark{a} & \citet{Pritchard12} \\
2007qd\tablenotemark{a} & \citet{McClelland10} \\
2007if\tablenotemark{a} & \citet{Yuan10} \\
2007it & \citet{Andrews11} \\
2007ru\tablenotemark{a} & \citet{Sahu09} \\
2007uy\tablenotemark{a} & \citet{Roy13} \\ 
2008D\tablenotemark{a} & \citet{Soderberg08} \\
2008fz & \citet{Drake10} \\
2008ge\tablenotemark{a} & \citet{Foley1008ge} \\
2008in & \citet{Roy11} \\
2009bw & \citet{Inserra1209bw} \\
2009dc\tablenotemark{a} & \citet{Yamanaka09a} \\
2009hd\tablenotemark{a} & \citet{EliasRosa11} \\
2009ig\tablenotemark{a} & \citet{Foley12c} \\
 & \citet{Marion13} \\
2009jf\tablenotemark{a} & \citet{Sahu11} \\
2009kf & \citet{Botticella10} \\
2009kn & \citet{Kankare12} \\ 
2009kr & \citet{Fraser10} \\
2009ku & \citet{Narayan11} \\
2009nr & \citet{Khan11} \\
2010aq & \citet{Gezari10} \\
2010ev & \cite{Gutierrez16} \\
2010jl\tablenotemark{a} & \citet{Fransson14} \\
 & \citet{Ofek14} \\
2010mb & \citet{BenAmi14} \\
PS1-10pm & \citet{McCrum15} \\
PS1-10afx & \citet{Chornock13} \\
PS1-10bzj & \citet{Lunnan13} \\
2011ay\tablenotemark{a} & \citet{Szalai15} \\
2011dh\tablenotemark{a} & \citet{Sahu13} \\
 & \citet{Marion14IIb} \\
{\bf 2011fe}\tablenotemark{a} & \citet{McClelland13} \\
 & \citet{Shappee13} \\
2011ht\tablenotemark{a} & \citet{Mauerhan13} \\
2011kl & \citet{Greiner15} \\
PTF11iqb\tablenotemark{a} & \citet{Smith15} \\
PS1-11ap & \citet{McCrum14} \\
2012ap\tablenotemark{a} & \citet{Liu1512ap} \\
2012au & \citet{Danmil13} \\
 & \citet{Takaki13} \\
2012bz\tablenotemark{a} & \citet{Schulze14} \\
2012dn\tablenotemark{a} & \citet{Chakradhari14} \\
2013ab & \citet{Bose15} \\
2013cq\tablenotemark{a} & \citet{Melandri14} \\
2013df & \citet{Morales14} \\
 & \citet{BenAmi15} \\
 & \citet{Maeda15} \\
2013ej\tablenotemark{a} & \citet{Huang15} \\
2013en & \citet{Liu1513en} \\
2013fu & \citet{Cano1413fu} \\
PS1-13arp & \citet{Gezari15} \\
iPTF13asv & \citet{Cao16} \\
2014J\tablenotemark{a} & \citet{Kawabata14} \\
 & \citet{Goobar14} \\
 & \citet{Vacca15} \\
 & \citet{Telesco15} \\
 & \citet{Marion15} \\
 & \citet{Jack15} \\
2015H & \citet{Magee16} \\
\enddata
\tablenotetext{a}{Some data for this object is publicly available.}\label{tab:missing}
\end{deluxetable}
\capstarttrue

\end{document}